# DFT exploration of pressure dependent physical properties of the recently discovered La$_3$Ni$_2$O$_7$ superconductor

Md. Enamul Haque, Ruman Ali, M. A. Masum, Jahid Hassan, S. H. Naqib*

Department of Physics, University of Rajshahi, Rajshahi 6205, Bangladesh

Corresponding author email: salehnaqib@yahoo.com

**Abstract**

The recent discovery of superconductivity in Ruddlesden-Popper bilayer nickelate La$_3$Ni$_2$O$_7$ under pressure has drawn a lot of interest. La$_3$Ni$_2$O$_7$ is isostructural with cuprates in some respect. Investigation of its properties will undoubtedly provide new insights into high-$T_c$ superconductivity. In the present work, we study structural, mechanical, elastic, optoelectronic, thermophysical properties, and Fermi surface topology of La$_3$Ni$_2$O$_7$ under pressure within the range of 30-40 GPa employing the density functional theory (DFT) based first-principles method. The calculated structural parameters agree well with the earlier experimental findings. The structural, mechanical, and thermodynamical stability is justified across the entire pressure range. The computed elastic moduli classify the compound as ductile, and the material's ductility is largely unaffected by pressure. The compound has a high level of machinability index and dry lubricity. Each anisotropy factor exhibits an elastically anisotropic nature. The electronic band structure reveals metallic feature of La$_3$Ni$_2$O$_7$ while the density of states at the Fermi level is roughly unchanged with increasing pressure. The Fermi surfaces reveal both hole and electron-like structures. The Debye temperature, thermal conductivity, and melting temperature increase with increasing pressure, but in an anomalous manner. The characteristic peaks in refractive index, reflectivity, and photoconductivity exhibit a small shift towards higher energy for all polarizations of the electric field vector with increasing pressure. The investigated material might be a good ultraviolet radiation absorber and can be used as an anti-reflection system. The effect of pressure on optical properties as well as optical anisotropy is weak. Moreover, the pressure dependent electronic density of states at the Fermi level, pressure induced negligible variations in the repulsive Coulomb pseudopotential, and the changes in the Debye temperature have been used to explore the effect of pressure on the superconducting transition temperature in this study.

## 1. Introduction

The discovery of superconductivity in nickelates marks a significant breakthrough in the field of condensed matter physics. In 1986, the discovery of superconducting copper oxides (cuprates) by Bednorz and Müller [1] sparked a quest to find analogous unconventional superconductivity in other material systems. In 1999, Anisimov *et al.* [2] proposed that nickelates could exhibit superconductivity, suggesting that nickel oxides might serve as direct analogs to cuprate superconductors. The investigation of transition metal oxides, especially those involving copper (cuprates) and nickel (nickelates), marked the beginning of the path toward nickelate superconductivity. In late 2019, superconductivity was first observed in the infinite-layer nickelate $Nd_{0.8}Sr_{0.2}NiO_2$, with superconducting transition temperature ($T_c$) ranging from 9 to 15 K [3,4]. This finding implied that nickelates might host electronic correlations akin to those of cuprates. Both nickelates and cuprates feature a layered perovskite structure, where the active layers are planes of nickel oxide ($NiO_2$) or copper oxide ($CuO_2$) respectively [5–8]. These layers are separated by rare-earth or alkaline-earth metals. Their electronic characteristics, especially superconductivity, are greatly influenced by these planes. Both families include partially filled $d$-orbitals of transition metal ions ($Cu^{2+}$ in cuprates and $Ni^2$ or $Ni^3$ in nickelates) [4,9,10]. This leads to strong electron correlations, which are essential for their unconventional superconductivity. Both cuprates and nickelates systems have nearly flat (non-dispersing) bands below the Fermi level [11–13]. These flat bands are essential ingredients for the dominant physics in cuprates and nickelates, potentially leading to all kinds of interesting physics like magnetic order, superconductivity, etc. Cuprates and nickelates have intriguing phase diagrams, particularly when considering superconductivity. The phase diagrams of both cuprate and nickelate are quite similar in that both have a domed shape superconducting region [10,14,15]. There are also dissimilarities in the phase diagram as well. These differences might offer a fundamentally novel perspective on the physics of superconductivity.

Lanthanum nickelate, or $La_3Ni_2O_7$, belongs to the Ruddlesden-Popper family of layered nickel oxides and is also the first bulk nickelate superconductor [16], known for its unique structural and electronic properties. $La_3Ni_2O_7$ crystallizes in rock-salt LaO layers and distorted vertex-sharing $NiO_6$ octahedra in orthorhombic symmetry [17]. The structure stacks along the $c$-direction and can be described as an intergrowth of a LaO fluorite-type layer and two $NiO_6$ octahedra planes. The layered structure allows for distinctive electron transport characteristics including high-temperature superconductivity under certain conditions of pressure. Hualei Sun *et al.* [15] discovered that $La_3Ni_2O_7$ demonstrates superconductivity under high pressure, reaching a maximum $T_c$ of 80 K at pressures ranging from 14.0 to 43.5 GPa, as measured using high-pressure resistance and mutual inductive magnetic susceptibility techniques. The nickel ions in $La_3Ni_2O_7$ have mixed oxidation states, which contribute to its complex electronic structure and magnetic interactions. These properties make $La_3Ni_2O_7$ a promising material for investigating unconventional superconductivity and the development of new electronic devices.

In this study, the compound $La_3Ni_2O_7$ is explored using the Kohn Sham density functional theory (KS-DFT) under various hydrostatic pressures. There are a few previous theoretical and experimental works on our chosen material, especially on electronic (including band structure, electronic density of states, and Fermi surface topology) and superconducting state properties [16,18–23]. As far as we know, the majority of the physical properties of these compounds, such as elastic, thermo-mechanical properties, and energy-dependent optical constants under various hydrostatic pressures have not been studied so far. For instance, there is currently no cited research on the elastic properties of these compounds, including Cauchy pressure, tetragonal shear modulus, hardness, anisotropy in elastic moduli, machinability index, Pugh's index, Kleinman parameter, Poisson's ratio, etc. Understanding the elastic properties and mechanical anisotropy is essential to fully realizing a compound's potential for various applications. Furthermore, the thermo-mechanical properties of this compound, including the Debye temperature, sound velocities, Grüneisen parameter, melting temperature, and others, have not been investigated. Investigating thermal properties provides insights into a compound's behavior under varying temperatures and pressures. Additionally, there is still a lack of theoretical insight into the optical properties of this material. Optical properties are important to know to select a material for optoelectronic device applications. The physical properties of solids can be adjusted by applying pressure [24,25]. Changes in the electronic ground state brought on by pressure are particularly interesting. For instance, high-$T_c$ cuprates exhibit considerable variations in their superconducting transition temperature, which are dependent on pressure and hole content [10,26,27].

Therefore, it is evident from the discussion above that many physical features of this compound remain unexplored. In this study, we seek to fill this research gap, which forms the main motivation for our current investigation. Therefore, the main theoretical aim of this paper is to comprehensively investigate the structural, elastic, mechanical, electronic, acoustic, optoelectronic, and thermal properties of the $La_3Ni_2O_7$ under varying pressure conditions. In addition to being a layered compound, $La_3Ni_2O_7$ has a lot of intriguing properties [28–30] that make it a material that requires careful consideration and investigation. The potential impact of pressure on superconductivity has been investigated. The results obtained are compared with those found in previous studies where available.

The rest of the paper is organized as follows: Section 2 briefly overviews the computational methods used in this study. Section 3 details the crystal structure and presents the computational results along with their analysis. Finally, Section 4 summarizes the key findings of the study.

## 2. Computational methodology

The most commonly used method for first-principles calculations of crystalline solids' structural and electrical properties is Density Functional Theory (DFT) [31,32], utilizing periodic Bloch boundary conditions. In this approach, the ground state properties of a compound are determined by solving the Kohn-Sham equation [33], which incorporates both exchange and correlation effects in the system's total energy. The accurate execution of the *ab-initio* computations depends in large part on the choice of the appropriate exchange-correlation scheme. We tested several exchange-correlation schemes (including GGA-PBESOL and LDA) to optimize the geometry of $La_3Ni_2O_7$ and found that the generalized gradient approximation using the Perdew-Burke-Ernzerhof (GGA-PBE) [34] functional provides the best results for the crystal structure. Therefore, we employed the GGA to model electron exchange correlations using the CAmbridge Serial Total Energy Package (CASTEP) [35], which is specifically designed for performing quantum mechanical DFT-based calculations. In general, GGA relaxes the electronic density distribution, making it more spatially extended than the more localized form of the density distribution used in the LDA (local density approximation) [36,37]. The electron-ion interactions were calculated using ultra-soft Vanderbilt-type pseudopotentials. This scheme relaxes the norm-conserving constraint, producing a smooth and computationally efficient pseudopotential that reduces computational time without significantly compromising accuracy.

The ground state of the crystal structure for $La_3Ni_2O_7$, with orthorhombic symmetry and space group *Fmmm* (No. 69) [8], was determined using the BFGS (Broyden-Fletcher-Goldfarb-Shanno) minimization technique [38]. The electronic orbitals of La, Ni, and O atoms are taken into account to form the valence and conduction states, respectively: [Xe] $5d^1$ $6s^2$, [Ar] $3d^8$ $4s^2$, and [He] $2s^2$ $2p^4$. The total energies of each crystal cell are calculated using periodic boundary conditions. The plane wave basis set is used to expand the trial wave functions. A cut-off energy of 400 eV was used for the plane wave expansion. The *k*-point sampling in the reciprocal space (Brillouin zone) for the compound was done using a dense mesh of 10×10×3 special points, employing the Monkhorst-Pack grid scheme [39]. The selections made for the *k*-point set and cut-off energy ensure a high degree of convergence in the computations of total energy versus cell volume. During geometry optimization, the structure is relaxed up to a convergence threshold for energy of $5\times10^{-6}$ eV-atom$^{-1}$, maximum force of 0.01 eV Å$^{-1}$, maximum stress of 0.02 GPa, maximum force of 0.01 eV Å$^{-1}$, and maximum displacement of $5\times10^{-4}$ Å.

The "stress-strain" method [40] in the CASTEP program were used to estimate elastic constants. An orthorhombic crystal has nine separate elastic constants ($C_{11}$, $C_{22}$, $C_{33}$, $C_{44}$, $C_{55}$, $C_{66}$, $C_{12}$, $C_{13}$, $C_{23}$). All elastic properties, such as the bulk modulus ($B$) and shear modulus ($G$), can be calculated from the values of the single crystal elastic constants $C_{ij}$ using the Voigt-Reuss-Hill (VRH) method [41,42]. The total density of states (TDOS), the partial density of states (PDOS), and the electronic band structure are obtained using the optimized geometry of $La_3Ni_2O_7$.

The complex dielectric function $\varepsilon(\omega) = \varepsilon_1(\omega) + i\varepsilon_2(\omega)$ is used to investigate the optical properties of $La_3Ni_2O_7$ compounds in the ground state. In the momentum representation of matrix elements between unoccupied and occupied electronic states in the conduction and valence states, the imaginary part, $\varepsilon_2(\omega)$, of the complex dielectric function was computed using the CASTEP-supported formula, which can be expressed as [43]:

$$\varepsilon_2(\omega) = \frac{2e^2\pi}{\Omega\varepsilon_0}\sum_{k,v,c}|\langle\psi_k^c|\hat{u}.\vec{r}|\psi_k^v\rangle|^2\,\delta(E_k^c - E_k^v - E) \qquad (1)$$

In this formula based on time-dependent perturbation theory, $e$ is the electronic charge, $\Omega$ is the unit cell's volume, $\omega$ is the angular frequency (or equivalently energy) of the incident electromagnetic wave (photon), and at a given wave vector $k$, $\psi_k^c$ and $\psi_k^v$ are the conduction and valence band wave functions, respectively. During the optical transition, the delta function forces energy conservation and momentum to be held. The Kramers-Kronig relationships are used to derive the real part [$\varepsilon_1(\omega)$] of the dielectric function from the imaginary part $\varepsilon_2(\omega)$. The degree of electric polarization is represented by $\varepsilon_1(\omega)$, and $\varepsilon_2(\omega)$ reflects the optical absorption of the material [44,45]. All optical parameters (the optical conductivity, refractive index, absorption coefficient, reflectivity, and energy loss function) can be computed from $\varepsilon_1(\omega)$, and $\varepsilon_2(\omega)$ as they are all connected. This procedure has been applied by several earlier studies to accurately compute all energy/frequency dependent optical parameters [43,46–48].

The following relationships [49] are used to calculate the real, $n(\omega)$, and imaginary, $k(\omega)$, components of the complex refractive index:

$$n(\omega) = \frac{1}{\sqrt{2}}[\{\varepsilon_1(\omega)^2 + \varepsilon_2(\omega)^2\}^{1/2} + \varepsilon_1(\omega)]^{1/2} \qquad (2)$$

$$k(\omega) = \frac{1}{\sqrt{2}}[\{\varepsilon_1(\omega)^2 + \varepsilon_2(\omega)^2\}^{1/2} - \varepsilon_1(\omega)]^{1/2} \qquad (3)$$

Again, by using the complex refractive index components, the reflectivity, or $R(\omega)$, can be computed as follows [49]:

$$R(\omega) = \left|\frac{\tilde{n}-1}{\tilde{n}+1}\right| = \frac{(n-1)^2 + k^2}{(n+1)^2 + k^2} \qquad (4)$$

Finally, the optical conductivity, $\sigma(\omega)$, the absorption coefficient, $\alpha(\omega)$, and the energy loss function, $L(\omega)$, can be obtained from the following equations [49]:

$$\alpha(\omega) = \frac{4\pi k(\omega)}{\lambda} \tag{5}$$

$$\sigma(\omega) = \frac{2W_{cv}\hbar\omega}{\vec{E}_0^2} \tag{6}$$

$$L(\omega) = Im\left(-\frac{1}{\varepsilon(\omega)}\right) \tag{7}$$

here, $W_{cv}$ in the above equation is the transition probability per unit time.

To investigate the pressure dependent elastic, thermodynamic, and structural properties, we used the energy-volume data obtained from the third-order Birch-Murnaghan equation [50].

## 3. Results and Discussions

### 3.1 Structural Properties and Cohesive Energy

The crystal structure of $La_3Ni_2O_7$ is orthorhombic with space group *Fmmm* (No. 69) [17,18,21,51]. The schematic diagram of the crystal structure of $La_3Ni_2O_7$ is shown in **Figure 1**. The unit cell of $La_3Ni_2O_7$ includes forty-eight atoms with twelve lanthanum eight nickel and twenty-eight oxygen atoms. The formula unit per unit cell is, therefore, 4.

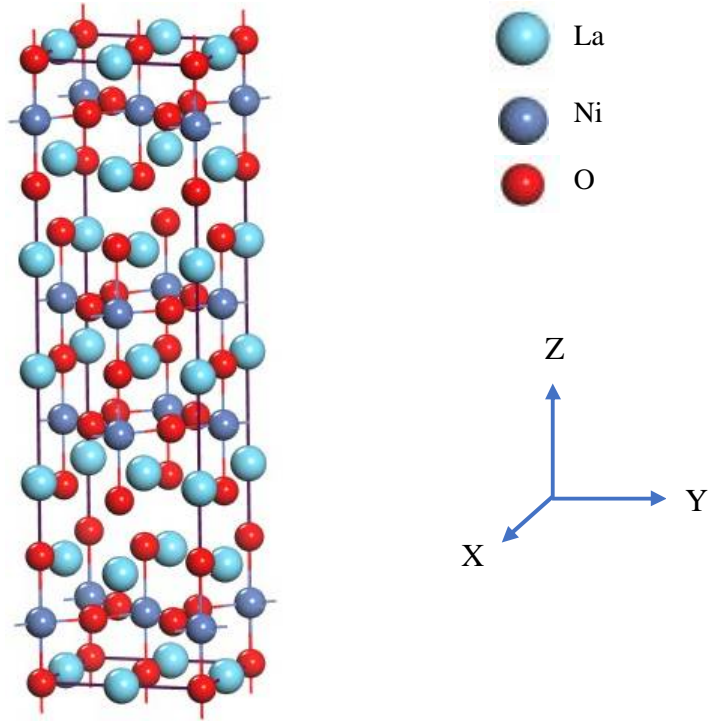

**Figure 1.** Three-dimensional schematic representation of the crystal structure of $La_3Ni_2O_7$ compound.

It is seen that La1, La2, Ni, and O1, O2, and O3 atoms occupied (0, 0, 0.321), (0, 0, 0.50), (0, 0, 0.096), (0.250, 0.250, 0.095), (0, 0, 0.204), and (0, 0, 0) positions respectively. **Table 1** lists the fractional coordinates for the La, Ni, and O atoms in $La_3Ni_2O_7$.

**Table 1.** The fractional coordinates of the atoms within the unit cell for $La_3Ni_2O_7$.

| Compound | Atoms | Fractional coordinates | | | Reference |
|---|---|---|---|---|---|
| | | X | y | z | |
| $La_3Ni_2O_7$ | Ni | 0 | 0 | 0.096 | [18] |
| | La1 | 0 | 0 | 0.321 | |
| | La2 | 0 | 0 | 0.50 | |
| | O1 | 0.250 | 0.250 | 0.095 | |
| | O2 | 0 | 0 | 0.204 | |
| | O3 | 0 | 0 | 0 | |

The results of first-principles calculations of ground state structural properties under different pressures including cohesive energy together with available experimental values are given in **Table 2**. Unfortunately, no experimental or theoretical values of the lattice parameters under high pressure except for 30 GPa are available for comparison. Pressure-dependent lattice parameters $a/a_0$, $b/b_0$, $c/c_0$, and $V/V_0$, (where $a_0$, $b_0$, $c_0$, and $V_0$ are the values of structure parameters and cell volume at 30 GPa) are plotted in **Figure 2**. The rate at which $c/c_0$ decreased was significantly greater than that of $a/a_0$ and $b/b_0$. Therefore, it can be concluded that the *c*-axis is easily compressed compared to the other two.

**Table 2.** The effect of pressure on the optimized lattice constants (*a*, *b*, and *c* in Å), cell volume *V* in Å$^3$, and cohesive energy ($E_{coh}$ in eV/atoms) of $La_3Ni_2O_7$.

| Pressure (GPa) | a | b | c | V | $E_{coh}$ | Reference |
|---|---|---|---|---|---|---|
| 30 | 5.238 | 5.215 | 19.793 | 540.669 | -5.38 | This work |
| 29.5* | 5.289 | 5.218 | 19.734 | 544.620 | -- | Expt. [18] |
| 32 | 5.227 | 5.204 | 19.742 | 537.008 | -5.24 | This work |
| 34 | 5.215 | 5.192 | 19.695 | 533.267 | -5.10 | This work |
| 36 | 5.204 | 5.181 | 19.648 | 529.748 | -4.96 | This work |
| 38 | 5.197 | 5.174 | 19.591 | 526.788 | -4.83 | This work |
| 40 | 5.187 | 5.163 | 19.546 | 523.451 | -4.69 | This work |

\* We use 29.5 GPa for comparison, as there are no experimental data available for 30 GPa.

We have determined the thermodynamic stability by computing the cohesive energy per atom, following the methodology used in Refs [26,27]. In addition, the cohesive energy, $E_{coh}$ has been calculated using the following equation:

$$E_{coh} = \frac{E_{La_3Ni_2O_7} - 3E_{La} - 2E_{Ni} - 7E_{O}}{12} \qquad (8)$$

where, $E_{La_3Ni_2O_7}$ is total energy per formula unit of La$_3$Ni$_2$O$_7$ and $E_{La}$, $E_{Ni}$ and $E_O$ are the total energies of single La, Ni, and O atoms in the solid state, respectively. As shown in **Table 2**, the values of cohesive energy per atom are negative, indicating that the La$_3$Ni$_2$O$_7$ structure is thermodynamically stable [28]. The cohesive energy varies with increasing pressure. We observe no abrupt changes in cohesive energy or structural parameters in the pressure range under consideration. This implies no signature of thermodynamic instability in this pressure range.

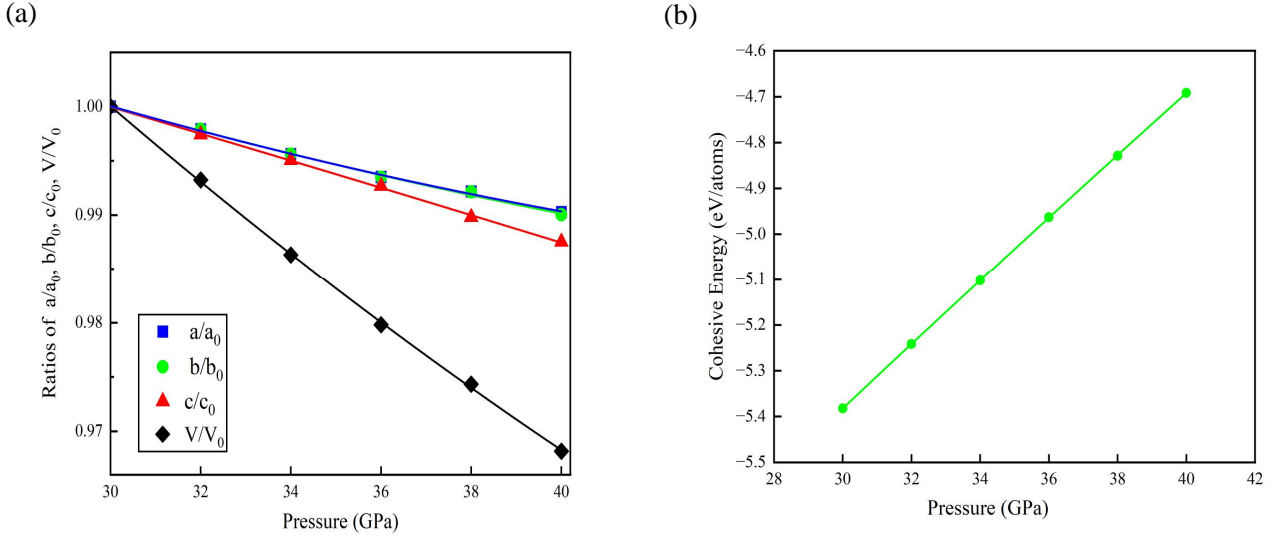

**Figure 2.** (a) Normalized parameters *a/a$_0$, b/b$_0$, c/c$_0$, V/V$_0$* and (b) cohesive energy/atom of La$_3$Ni$_2$O$_7$ under different pressures.

## 3.2 Mechanical and Elastic Properties

Elastic properties play a crucial role in materials science and technology. These properties link the mechanical and dynamical behavior of crystals and give important information concerning the nature of the forces operating in solids. In particular, they provide information on the stability and stiffness of materials [54,55]. There are no theoretical or experimental results available to compare with the present results. Thus, our results should serve as a reference for future investigations.

Nine independent strains are necessary to calculate the elastic constants of orthorhombic La$_3$Ni$_2$O$_7$ compound. The elastic constants calculated for different pressures are shown in **Table 2**. For the orthorhombic crystals, the mechanical stability criterion under hydrostatic pressure can be written as [56]:

$$\tilde{C}_{11} + \tilde{C}_{22} - 2\tilde{C}_{12} > 0,$$
$$\tilde{C}_{11} + \tilde{C}_{33} - 2\tilde{C}_{13} > 0,$$
$$\tilde{C}_{22} + \tilde{C}_{33} - 2\tilde{C}_{23} > 0,$$
$$\tilde{C}_{ii} > 0 \ (i = 1\sim6),$$
$$\tilde{C}_{11} + \tilde{C}_{22} + \tilde{C}_{33} + 2\tilde{C}_{12} + 2\tilde{C}_{13} + 2\tilde{C}_{23} > 0,$$

where $\tilde{C}_{ii} = C_{ii} - P$ ($i = 1\sim6$), $\tilde{C}_{12} = C_{12} + P$, $\tilde{C}_{13} = C_{13} + P$, $\tilde{C}_{23} = C_{23} + P$. The elastic constants under different pressures obey these stability criteria, implying that the orthorhombic $La_3Ni_2O_7$ is mechanically stable within 30 - 40 GPa. As seen from **Table 3** [cf. **Figure 3** (a)], almost all the elastic constants increased roughly monotonically with pressure except a few at a certain pressure, which might be a sign of structural instability. Sign of structural instability was also found in previous work [16]. This is also an indication that the strength and bonding nature are complex functions of pressure. $C_{11}$, $C_{22}$, and $C_{33}$ correspond to the resistance to linear compression in the [100], [010], and [001] directions, respectively while the remaining elastic constants are primarily related to the elasticity of shape. In the entire pressure range of our calculations, the values of $C_{11}$, $C_{22}$, and $C_{33}$ were significantly greater than those of the other elastic constants. $C_{33}$ is quite low compared to $C_{11}$, and $C_{22}$ except $C_{33}$ at 30 GPa. This indicates that the crystal exhibits the highest compressibility under uniaxial strain when the stress is applied along the *c*-axis. Such anisotropic mechanical behaviors are associated with the anisotropic nature of underlying chemical bondings and the layered character of the crystal structure.

**Table 3.** Single-crystal elastic constants, $C_{ij}$, tetragonal shear modulus, $C'$, for $La_3Ni_2O_7$ under various pressures (all in GPa).

| Pressure | $C_{11}$ | $C_{22}$ | $C_{33}$ | $C_{44}$ | $C_{55}$ | $C_{66}$ | $C_{12}$ | $C_{13}$ | $C_{23}$ | $C'$ |
|---|---|---|---|---|---|---|---|---|---|---|
| 30 | 414.42 | 412.86 | 506.37 | 92.16 | 92.18 | 111.83 | 214.90 | 205.30 | 204.41 | 99.76 |
| 32 | 441.42 | 442.51 | 429.25 | 95.38 | 93.63 | 170.53 | 238.04 | 214.36 | 215.66 | 101.69 |
| 34 | 461.03 | 454.58 | 443.08 | 98.60 | 98.65 | 174.86 | 248.60 | 224.18 | 220.68 | 106.22 |
| 36 | 444.90 | 445.30 | 364.33 | 97.03 | 96.99 | 128.05 | 230.67 | 233.71 | 233.96 | 107.12 |
| 38 | 509.70 | 508.99 | 402.36 | 109.56 | 109.89 | 138.04 | 294.02 | 266.53 | 266.13 | 107.84 |
| 40 | 506.37 | 506.26 | 483.89 | 111.33 | 111.33 | 190.39 | 286.14 | 244.54 | 244.47 | 110.12 |

The elastic constant $C_{44}$ measures the resistance to shear deformation caused by tangential stress applied to the (100) plane along the [010] direction of the compound. It can be observed that for $La_3Ni_2O_7$, the value of $C_{44}$ is significantly lower than $C_{11}$, $C_{22}$, and $C_{33}$ suggesting that the compound is more easily deformed by shear along any of the three crystallographic directions than it does unidirectional compression along those directions. **Table 3** shows that this condition is consistently met, suggesting that the compound under investigation is susceptible to shear deformations. Thus, shearing strain is expected to control the mechanical failure of $La_3Ni_2O_7$. The off-diagonal shear components, or $C_{12}$ and $C_{13}$, are two of the other elastic constants that have been recorded. They relate to the resistance the compound provides against various types of shape distortions. For $La_3Ni_2O_7$, $C_{44}$ is lower than $C_{66}$, indicating that the shear along the (100) plane is easier than the shear along the (001) plane. Since $(C_{11}+C_{12}) > C_{33}$, the bonding in the (001) plane is more elastically rigid than that along the *c*-axis, and the elastic tensile modulus is higher on the (001) plane than it is along the *c*-axis. The tetragonal shear modulus, $C'$ [given by, $0.5(C_{11} - C_{12})$] is another useful elastic parameter that measures the crystal's stiffness (the resistance to shear deformation by shear stress applied in the (110) plane in the [1$\bar{1}$0] direction). The tetragonal shear modulus is also referred to as the shear constant. The degree of dynamic stability of a crystal is linked with the value of $C'$. A positive shear constant in a material indicates stability against tetragonal distortions; otherwise, the material is likely to be dynamically unstable. The calculated value for $La_3Ni_2O_7$ is provided in **Table 3**. It is observed that the shear constant of $La_3Ni_2O_7$ is positive at all pressures, indicating that the compound is expected to be dynamically stable.

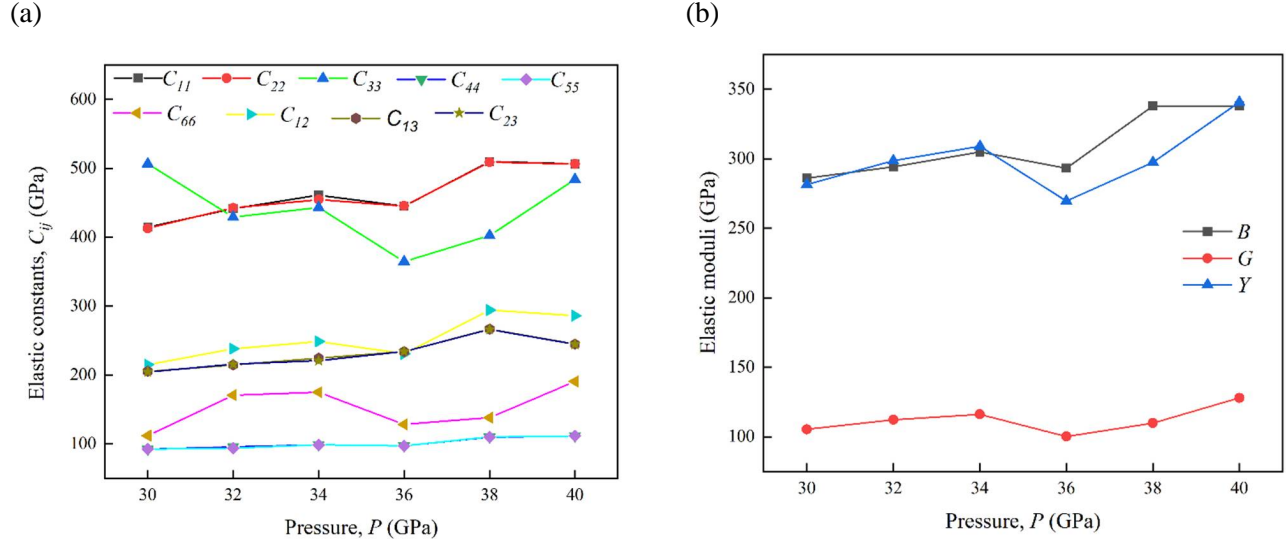

**Figure 3.** (a) Single-crystal elastic constant, and (b) Elastic moduli under various pressures.

The single crystal elastic constants, $C_{ij}'s$, allow for the determination of different bulk elastic moduli, anisotropy indicators, and Poisson's ratio. Using Voigt's approximation [57], the isotropic bulk and shear moduli can be calculated by taking a linear combination of various elastic constants. The shear and bulk moduli approximated by Voigt are denoted as $G_V$ and $B_V$, respectively. However, using the single crystal elastic constants, Ruess [34] provides an alternative estimate for the isotropic bulk and shear moduli. The shear and bulk moduli approximated by Ruess are denoted as $B_R$ and $G_R$, respectively. Hill [58] later showed that the values approximated by Ruess represent the lower limit for the polycrystalline elastic moduli, while those approximated by Voigt represent the upper limit. The real value lies between the Voigt and Reuss bounds. According to Hill, the arithmetic means for the bulk modulus and shear modulus are more representative of the actual situation. The polycrystalline values of bulk modulus ($B$), shear modulus ($G$), Young's modulus ($Y$), and Poisson's ratio ($v$) are calculated by using the following equations [59–61]:

$$B = \frac{B_V + B_R}{2} \quad (9)$$

$$G = \frac{G_V + G_R}{2} \quad (10)$$

$$Y = \frac{9BG}{3B + G} \quad (11)$$

$$v = \frac{3B - 2G}{2(3B + G)} \quad (12)$$

**Table 4** presents the values of polycrystalline elastic moduli (*B*, *G*, and *Y*), which characterize the bulk elastic response of a solid, along with Cauchy pressure (*CP*), Pugh's ratio (*G/B*), and Poisson's ratio (*v*) at different pressures. Additionally, **Figure 4** illustrates the variations in bulk, shear, and Young's moduli, Cauchy pressure, Pugh's ratio, and Poisson's ratio for pressures ranging from 30 GPa to 40 GPa. This study primarily focuses on examining how pressure influences the various properties of the compound $La_3Ni_2O_7$.

The isotropic shear modulus and bulk modulus are overall indicators of a material's hardness. The bulk modulus measures resistance to volume change brought on by applied pressure while the shear modulus measures resistance to reversible deformations brought on by shearing stress. A large shear modulus value in a solid denotes strong directional bonding between atoms [62]. The covalent character of a material increases as the value of the Young's modulus increases [63]. Young's modulus measures the stiffness (resistance) of an elastic material to changes in its length and serves as an indicator of thermal shock resistance [64,65]. The calculated Young's modulus (*Y*) of the nickelate is relatively high, suggesting that $La_3Ni_2O_7$ is expected to be a fairly stiff material. Nonmonotonic variation in elastic moduli with pressure (**Table 4** and **Figure 3** (b)) once again demonstrates the complex nature of the atomic bonding characteristic of $La_3Ni_2O_7$. However, among all the three moduli, the shear modulus is the lowest at all pressures. This suggests that the elastic failure of $La_3Ni_2O_7$ should be controlled by the shape-changing strain. A number of thermophysical factors are linked to elastic moduli. For example, the lattice thermal conductivity ($K_L$) and Young's modulus of a solid are related as $K_L \sim \sqrt{Y}$ [66].

**Table 4.** The bulk modulus *B*, shear modulus *G* Young's modulus *Y*, Cauchy pressure *CP* (all in GPa), Pugh's ratio *G/B*, and Poisson's ratio *v* for the compound $La_3Ni_2O_7$ under different pressures.

| Pressure | $B_V$ | $B_R$ | $B$ | $G_V$ | $G_R$ | $G$ | $G/B$ | $Y$ | $N$ | $CP$ |
|---|---|---|---|---|---|---|---|---|---|---|
| 30 | 286.98 | 285.13 | 286.06 | 106.50 | 104.05 | 105.28 | 0.37 | 281.33 | 0.34 | 122.74 |
| 32 | 294.37 | 293.95 | 294.16 | 114.92 | 109.43 | 112.17 | 0.38 | 298.57 | 0.33 | 142.66 |
| 34 | 305.07 | 304.51 | 304.79 | 118.77 | 113.35 | 116.06 | 0.38 | 308.97 | 0.33 | 150.00 |
| 36 | 294.58 | 291.72 | 293.15 | 101.49 | 98.68 | 100.09 | 0.34 | 269.58 | 0.35 | 133.64 |
| 38 | 341.60 | 334.17 | 337.89 | 111.12 | 108.55 | 109.84 | 0.33 | 297.30 | 0.35 | 184.46 |
| 40 | 338.53 | 337.37 | 337.95 | 130.70 | 125.01 | 127.86 | 0.38 | 340.61 | 0.33 | 174.81 |

Pugh suggested that the *G/B* ratio also known as Pugh's ratio, represents a solid's brittleness or ductility [67–69]. A material is considered ductile if the *B/G* ratio is below 0.57; if it is higher, the material is likely to be brittle. Furthermore, the *B/G* ratio illustrates the relative directionality of the bonding occurring within the material [70]. **Figure 4** (a) shows how pressure affects the Pugh's ratio of the $La_3Ni_2O_7$ material. It is seen from **Table 4** [cf. **Figure 4** (a)] that, the *G/B* of $La_3Ni_2O_7$ is less than 0.57 for all the pressures indicating that the compound is expected to be ductile. It is also obvious from the figure that pressure has little effect on the ductility of the material in this pressure range.

Poisson's ratio (*v*) is an essential parameter for examining various properties of a compound, including its brittle or ductile nature, compressibility, and bonding characteristics. The Poisson's ratio of a solid can range from -1.0 (when $B \ll G$) to 0.50 (when $B \gg G$) [71]. The critical value of the Poisson's ratio is 0.26 [61]. The material should behave ductile if the Poisson's ratio is greater than the critical value; on the other hand, if the value is lower, the material should be brittle. The nature of the interatomic forces in solids is related to the Poisson's ratio [57]. It is also established that central force interactions primarily govern the bonding in a solid when Poisson's ratio falls within the range of 0.25 to 0.50 [72,73]. However, if the ratio falls outside this range, non-central force interactions dominate the solid's bonding. In addition, Poisson's ratio provides insights into the presence of covalent and ionic bonding in a compound. For instance, the Poisson's ratio is generally around 0.25 for ionic materials and 0.10 for covalent materials [74]. According to **Table 4**, since the compound's Poisson's ratio, under the influence of pressure, falls within the specified range, it can be inferred that central forces dominate the bonding in the compound. The effect of pressure on the Poisson's ratio of the $La_3Ni_2O_7$ compound is shown in **Figure 4** (b). $La_3Ni_2O_7$ has Poisson's ratio greater than 0.26 under pressure, which indicates that it is ductile. Hence, we can obtain the same conclusion from both $G/B$ and $v$.

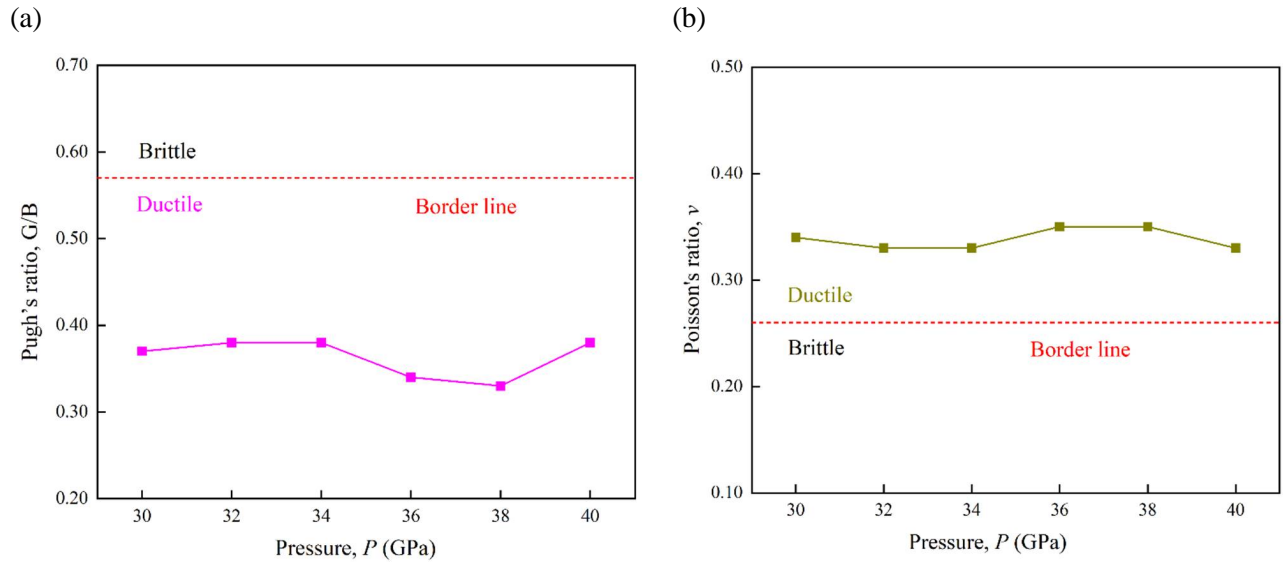

**Figure 4.** (a) Pugh's ratio and (b) Poisson's ratio of $La_3Ni_2O_7$ under varying pressure.

Another method to ascertain the brittleness or ductility of a material is by computing the Cauchy pressure, which can be expressed as $CP = (C_{12} - C_{44})$. The material should be ductile if the Cauchy pressure is positive; otherwise, it should be brittle [75]. The Cauchy pressure can also be used to describe the angular characteristic of atomic bonding in a material [76]. The existence of ionic bonding and covalent bonding in a material are associated with positive and negative values of the Cauchy pressure, respectively. It is seen from **Table 4** that the Cauchy pressure of $La_3Ni_2O_7$ is positive for all the pressures considered suggesting that the compound should be ductile in nature. It is also evident from **Table 4** that Cauchy pressure varies with applied pressures and the ductility of the $La_3Ni_2O_7$ material rises generally with increasing pressure. Pettifor's rule [76] suggests that materials with high positive Cauchy pressures tend to have more metallic bonds, making them more ductile. Conversely, materials with highly negative Cauchy pressures are characterized by more angular bonds, which makes them more brittle. As a result, the positive value of Cauchy pressure predicts that the $La_3Ni_2O_7$ compound is ductile in nature and metallic bonds are present. The results from Pugh's ratio, Poisson's ratio, and Cauchy pressure are entirely consistent with each other.

Other significant mechanical performance indicators, including the machinability index ($\mu^M$), Kleinman parameter ($\zeta$), and Vickers hardness ($H_V$), are computed using the commonly used formulas [77,78] and are provided in **Table 5**

$$\mu^M = \frac{B}{C_{44}} \tag{13}$$

$$\zeta = \frac{C_{11} + 8C_{12}}{7C_{11} + 2C_{12}} \tag{14}$$

$$H_V = \frac{(1-2\nu)Y}{6(1+\nu)} \tag{15}$$

**Table 5.** The calculated value of Vickers hardness, $H_V$, machinability index ($\mu^M$), Kleinman parameter ($\zeta$), and the effect of pressure on La$_3$Ni$_2$O$_7$ compound.

| Pressure (GPa) | Vickers Hardness, $H_V$ (GPa) | Machinability index, $\mu^M$ | Kleinman parameter, $\zeta$ |
|---|---|---|---|
| 30 | 11.20 | 3.104 | 0.641 |
| 32 | 12.72 | 3.084 | 0.658 |
| 34 | 13.16 | 3.091 | 0.658 |
| 36 | 9.98 | 3.021 | 0.641 |
| 38 | 11.01 | 3.084 | 0.689 |
| 40 | 14.51 | 3.036 | 0.679 |

The Machinability Index is a critical performance measure for potential engineering applications. It provides an indication of how easily a solid can be cut or shaped in various ways. A higher $\mu^M$ value signifies easier shaping of the material and enhanced dry lubrication [79,80]. These solids can be quickly shaped into the desired geometry and require less power to machine. The machinability index of La$_3$Ni$_2$O$_7$ at various pressures has been calculated and is listed in **Table 5**. The machinability index of La$_3$Ni$_2$O$_7$ is high at all pressures. Overall, $\mu^M$ discloses non-monotonic pressure dependence although the values are roughly equal. It is worth mentioning that the element aluminum (Al), which is soft, ductile, and machinable, has a $\mu^M$ value of approximately 2.0 [78]. All these $\mu^M$ values suggest a high level of machinability of La$_3$Ni$_2$O$_7$, similar to larger than that of many technologically significant MAX phase nanolaminates and related compounds [46,47,79,81].

The Kleinman parameter is an elastic property that describes the relative positions of the cation and anion sublattices during volume-conserving distortions, where the atomic positions are not determined by the crystal symmetry [82]. The Kleinman parameter ($\zeta$), also known as the internal strain parameter, measures how stable a compound is against distortions of the stretching and bending types. It is a unitless parameter [28]. The value of $\zeta$ ranges from zero to one ($0 \leq \zeta \leq 1$) [61]. According to Kleinman, the lower limit of $\zeta$ reflects a major contribution to resistance from bond stretching or contracting under external stress. The upper limit indicates a dominant contribution from bond bending in response to external stress. As seen from **Table 5** the calculated value of $\zeta$ of La$_3$Ni$_2$O$_7$ is greater than 0.6 for all the pressures considered. Therefore, it can be inferred that mechanical strengths in La$_3$Ni$_2$O$_7$ are primarily generated from the bond-bending contributions.

To fully understand a material's elastic and plastic properties, it is also essential to consider its hardness value. The estimated value of hardness of La$_3$Ni$_2$O$_7$ under various pressures is recorded in **Table 5**. This suggests that the compound being investigated has a reasonable level of hardness, comparable to some binary intermetallic compounds [83]. Overall, La$_3$Ni$_2$O$_7$ is hard, ductile, and highly machinable. Such a combination of mechanical properties makes it very suitable for engineering applications.

### 3.3 Elastic Anisotropy

Almost all the known crystalline materials are anisotropic both mechanically and elastically. Anisotropy in the elastic parameters of a crystalline solid explains the direction dependence of its mechanical properties. The nature and extent of elastic anisotropy are closely related to a number of physical processes, such as the formation of micro-cracks in solids, the motion of cracks, the development of plastic deformations in crystals, etc., which makes the study of anisotropic elastic properties of materials valuable. For example, the degree of anisotropy in the bonding strength for atoms positioned in different planes is measured by the shear anisotropy factors. The fields of applied engineering sciences and crystal physics would both benefit greatly from a thorough explanation of these properties. Therefore, it is essential to calculate the elastic anisotropy indices of La$_3$Ni$_2$O$_7$ in detail to understand their durability and possible applications under different types of external loading.

The degree of anisotropy in the bonding between atoms in various planes can be measured using the shear anisotropic factors. The shear anisotropy for an orthorhombic crystal can be quantified by following indicators [62,84]:

The Zener anisotropy factors (*A*) are calculated using the following formula:

$$A = \frac{2C_{44}}{C_{11} - C_{12}} \qquad (16)$$

The shear anisotropy factor for {100} shear planes between the ⟨011⟩ and ⟨010⟩ directions is given by:

$$A_1 = \frac{4C_{44}}{C_{11} + C_{33} - 2C_{13}} \qquad (17)$$

The shear anisotropy factor for the {010} shear plane between ⟨101⟩ and ⟨001⟩ directions is given by:

$$A_2 = \frac{4C_{55}}{C_{22} + C_{33} - 2C_{23}} \qquad (18)$$

and the shear anisotropy factor for the {001} shear planes between ⟨110⟩ and ⟨010⟩ directions is given by:

$$A_3 = \frac{4C_{66}}{C_{11} + C_{22} - 2C_{12}} \qquad (19)$$

The Zener anisotropy factors, $A$, and the shear anisotropy factors $A_1$, $A_2$, and $A_3$ must be one for an isotropic crystal, while any other value, smaller or greater than one, measures the degree of elastic anisotropy [85]. All the calculated values of various anisotropy factors under pressure from 30 GPa to 40 GPa for $La_3Ni_2O_7$ are recorded in **Table 6** [cf. **Figure 5** (a)]. As shown in **Table 6**, the calculated values of $A$, $A_1$, $A_2$ and, $A_3$ are different from unity, indicating that the compound $La_3Ni_2O_7$ exhibits anisotropy in response to shearing stress across different crystal planes. When the applied pressure increased from 30–40 GPa, $A_1$, $A_2$, and $A_3$ increased non-monotonically with pressure. The shear anisotropy factors $A_3$ are quite higher than $A_1$, and $A_2$. It is also evident from **Figure 5** (a) that pressure had a similar influence on the shear anisotropy of $\{100\}$ shear planes between the $\langle 011 \rangle$ and $\langle 010 \rangle$ directions, as well as the $\{010\}$ shear plane between $\langle 101 \rangle$ and $\langle 001 \rangle$ directions. Altogether, it can be asserted that the compound $La_3Ni_2O_7$ possesses moderately anisotropic behavior.

The universal anisotropy index ($A^U$, $d_E$), equivalent Zener anisotropy measure $A^{eq}$, anisotropy in compressibility $A_B$ and anisotropy in shear $A_G$ for crystals with any symmetry are calculated using the following formulas [86–89]:

$$A^U = \frac{B_V}{B_R} + 5\frac{G_V}{G_R} - 6 \geq 0 \tag{20}$$

$$d_E = \sqrt{A^U + 6} \tag{21}$$

$$A^{eq} = \left(1 + \frac{5}{12}A^U\right) + \sqrt{\left(1 + \frac{5}{12}A^U\right)^2 - 1} \tag{22}$$

$$A_B = \frac{B_V - B_R}{B_V + B_R} \tag{23}$$

$$A_G = \frac{G_V - G_R}{G_V + G_R} \tag{24}$$

Ranganathan and Ostoja Starzewski [87] proposed the concept of a universal anisotropy index ($A^U$), which provides a singular measure of anisotropy irrespective of the crystal symmetry. In contrast to all other anisotropy indicators, $A^U$ considers the influence of the bulk to the anisotropy of a solid for the very first time. A higher degree of elastic anisotropy would be indicated by a larger fractional difference between the Voigt and Reuss estimated bulk or shear modulus, according to the **Equation 20**. The values of $G_V/G_R$ and $B_V/B_R$ for $La_3Ni_2O_7$ show that $G_V/G_R$ has a slightly greater effect on $A^U$ than $B_V/B_R$. The universal anisotropy factor can only take two possible values: zero or positive. A value of $A^U$ equal to zero indicates that the crystal is isotropic, while any deviation from this value indicates the presence of anisotropy. The $A^U$ value for $La_3Ni_2O_7$, listed in **Table 6**, deviates from zero, indicating that the compound exhibits anisotropy in its elastic and mechanical properties within the given pressure range.

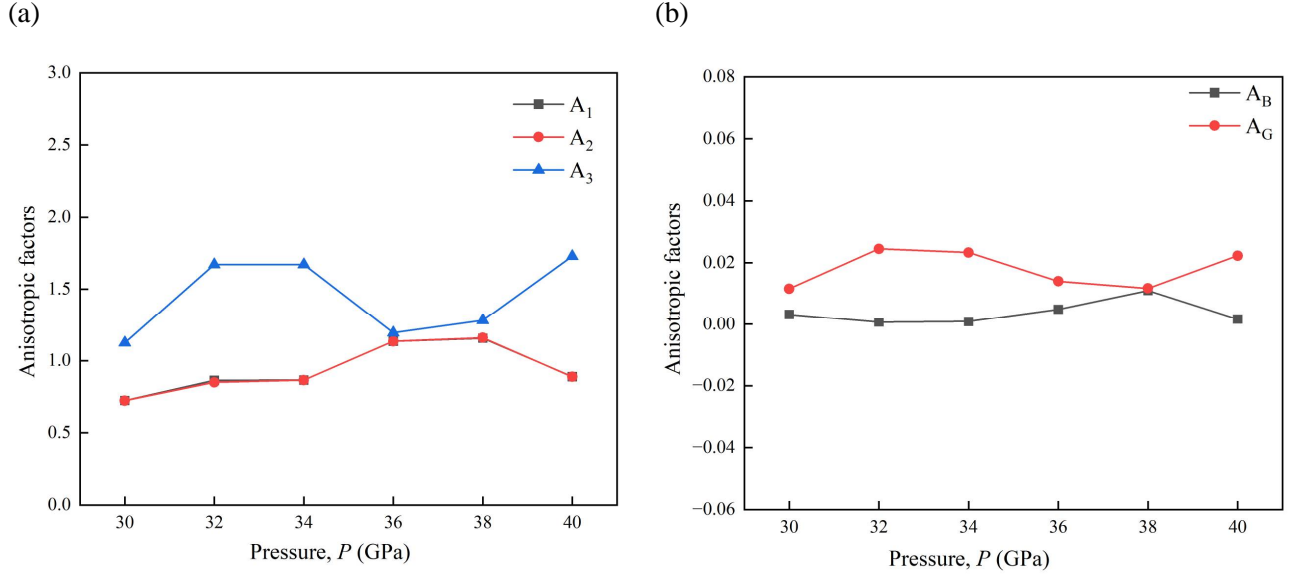

**Figure 5.** Anisotropy factors (a) $A_1$, $A_2$, and $A_3$ and (b) $A_B$ and $A_G$ of La$_3$Ni$_2$O$_7$ as a function of pressure.

The universal log-Euclidean index is given by the following expression [86,90]:

$$A^L = \sqrt{\left[\ln\left(\frac{B_V}{B_R}\right)\right]^2 + 5\left[\ln\left(\frac{C_{44}^V}{C_{44}^R}\right)\right]^2} \quad (25)$$

where,

$$C_{44}^R = \frac{5}{3}\frac{C_{44}(C_{11} - C_{12})}{3(C_{11} - C_{12}) + 4C_{44}} \quad \text{is the Reuss value of } C_{44}$$

and,

$$C_{44}^V = C_{44}^R + \frac{3}{5}\frac{(C_{11} - C_{12} - 2C_{44})^2}{3(C_{11} - C_{12}) + 4C_{44}} \quad \text{is the Voigt value of } C_{44}$$

Kube and Jong [86] mentioned that 90% of the inorganic crystalline compounds have an $A^L$ value of less than 1, and that the values of these compounds fall within the range of $0 \leq A^L \leq 10.26$. For a perfectly isotropic crystal, it is noted that $A^L = 0$. While it is difficult to determine whether a solid is layered solely based on the $A^L$ value, most (78%) of these high $A^L$ value inorganic crystalline compounds show layered structural characteristics [86,91]. From the discussions above, it is possible to predict the high anisotropy and layered nature of the titled compound. It is evident from **Table 6** that the value $A^L$ in non-zero so the compound La$_3$Ni$_2$O$_7$ the compound possesses anisotropy in elastic/mechanical properties in the considered pressure range. In our case, it is difficult to predict the value of $A^L$ whether the compound shows layered or non-layered structures.

In the case of an isotropic crystalline material, the value of $A^{eq}$ is 1 [61]. The calculated $A^{eq}$ value for La$_3$Ni$_2$O$_7$ is greater than 1 at all considered pressures, indicating that the compound is anisotropic. In an isotropic crystal, the values of $A_G$ and $A_B$ are both zero. While any deviation from zero (positive) indicates the extent of anisotropy. The pressure dependence of $A_G$ and $A_B$ for La$_3$Ni$_2$O$_7$ is presented in **Figure 5** (b). As can be seen, the value of $A_G$ was larger than that of $A_B$, and the value of $A_B$, was closer to zero over the whole pressure range investigated, implying that La$_3$Ni$_2$O$_7$ is largely isotropic in volume strain and slightly anisotropic in shear strain.

**Table 6.** Elastic anisotropy indices and the effect of pressure on the La$_3$Ni$_2$O$_7$ compound.

| Pressure (GPa) | $A$ | $A_1$ | $A_2$ | $A_3$ | $A_B$ | $A_G$ | $A^U$ | $d_E$ | $A^{eq}$ | $A^L$ | $A_{B_a}$ | $A_{B_c}$ |
|---|---|---|---|---|---|---|---|---|---|---|---|---|
| 30 | 0.924 | 0.723 | 0.722 | 1.125 | 0.0032 | 0.0116 | 0.124 | 2.47 | 1.38 | 0.0126 | 1.011 | 1.385 |
| 32 | 0.938 | 0.863 | 0.850 | 1.672 | 0.0007 | 0.0245 | 0.252 | 2.50 | 1.58 | 0.0072 | 0.987 | 0.852 |
| 34 | 0.928 | 0.865 | 0.865 | 1.672 | 0.0009 | 0.0233 | 0.241 | 2.50 | 1.56 | 0.0097 | 1.052 | 0.865 |
| 36 | 0.906 | 1.135 | 1.135 | 1.194 | 0.0049 | 0.0140 | 0.152 | 2.48 | 1.43 | 0.0194 | 0.996 | 0.626 |
| 38 | 1.016 | 1.156 | 1.160 | 1.282 | 0.0110 | 0.0117 | 0.141 | 2.48 | 1.41 | 0.0220 | 1.007 | 0.504 |
| 40 | 1.011 | 0.889 | 0.888 | 1.729 | 0.0017 | 0.0223 | 0.231 | 2.50 | 1.55 | 0.0034 | 1.001 | 0.789 |

In contrast to a cubic crystal, a crystal's bulk modulus depends on direction, and elastic anisotropy results from both the directional variation of the bulk modulus and shear anisotropy. Therefore, elastic anisotropy cannot be adequately described by the shear anisotropy factors alone. A solid's bulk modulus along different crystallographic axes can be easily determined using pressure-dependent lattice parameters or through single crystal elastic constants. The directional bulk modulus, which is represented by $B_a$, $B_b$ and $B_c$ along the *a*-, *b*- and *c*-axis, respectively, the isotropic bulk modulus ($B_{relax}$), and the anisotropies of the bulk modulus along the *a*- and *c*-axis ($A_{Ba}$ and $A_{Bc}$) of the La$_3$Ni$_2$O$_7$ compound are defined as follows [29,84]:

$$B_a = a\frac{dP}{da} = \frac{\Lambda}{1+\alpha+\beta} \qquad (26)$$

$$B_b = b\frac{dP}{db} = \frac{B_a}{\alpha} \qquad (27)$$

$$B_c = c\frac{dP}{dc} = \frac{B_a}{\beta} \qquad (28)$$

$$B_{relax} = \frac{\Lambda}{(1+\alpha+\beta)^2} \qquad (29)$$

where,

$$\Lambda = C_{11}+2C_{12}\alpha+C_{22}\alpha^2+2C_{13}\beta+C_{33}\beta^2+2C_{23}\alpha\beta$$

and,

$$\alpha = \frac{\{(C_{11} - C_{12})(C_{33} - C_{13})\} - \{(C_{23} - C_{13})(C_{11} - C_{13})\}}{\{(C_{33} - C_{13})(C_{22} - C_{12})\} - \{(C_{13} - C_{23})(C_{12} - C_{23})\}}$$

$$\beta = \frac{\{(C_{22} - C_{12})(C_{11} - C_{13})\} - \{(C_{11} - C_{12})(C_{23} - C_{12})\}}{\{(C_{22} - C_{12})(C_{33} - C_{13})\} - \{(C_{12} - C_{23})(C_{13} - C_{23})\}} b$$

$$A_{B_a} = \frac{B_a}{B_b} = \alpha \tag{30}$$

$$A_{B_c} = \frac{B_c}{B_b} = \frac{\alpha}{\beta} \tag{31}$$

The parameters derived from these relationships for the La$_3$Ni$_2$O$_7$ system are listed in **Table 7**, illustrating the effect of pressure on the material. The single crystal isotropic bulk modulus, or $B_{relax}$, is nearly identical to the value derived from the Reuss approximation. $\alpha$ and $\beta$ are the relative change of the *b* and *c* axis as a function of the deformation of the *a*- axis [29]. The compound under investigation is highly compressible when stress is applied along the *c*-direction, as indicated by the small value of $B_c$. Although there is a sudden increase at 30 GPa which indicates a sign of structural instability arising in the optimized crystal system. The material appears to have significant bonding anisotropy based on the varied values of $B_a$, $B_b$, and $B_c$. In comparison to the one within the *ab*-plane, the anisotropy is strongest along the *c*-direction. The compressibility anisotropies of the bulk modulus along the *a*-axis and *c*-axis with respect to the *b*-axis are $A_{B_a}$ and $A_{B_c}$, respectively. These factors have a unit value that represents the isotropic elastic behavior of crystals; a value that deviates from unity represents the degree of elastic anisotropy. It is seen **Table 7** that the anisotropy of linear bulk modulus along the *c*-axis is more significant than that along the *a*-axis.

**Table 7.** The bulk modulus ($B_{relax}$ in GPa), bulk modulus along the *a*-, *b*- and *c*-axis ($B_a$, $B_b$, $B_c$ in GPa), $\alpha$, and $\beta$ and anisotropies of the bulk modulus along the *a*- and *c*-axis ($A_{Ba}$ and $A_{Bc}$) of the compound La$_3$Ni$_2$O$_7$ and the effect of pressure on it.

| Pressure (GPa) | $B_{relax}$ | $B_a$ | $B_b$ | $B_c$ | $\alpha$ | $\beta$ | $A_{B_a}$ | $A_{B_c}$ |
|---|---|---|---|---|---|---|---|---|
| 30 | 285.14 | 781.56 | 773.05 | 1070.63 | 1.011 | 0.730 | 1.011 | 1.385 |
| 32 | 293.95 | 924.76 | 936.94 | 797.89 | 0.987 | 1.159 | 0.987 | 0.852 |
| 34 | 304.51 | 995.15 | 945.96 | 818.38 | 1.052 | 1.216 | 1.052 | 0.865 |
| 36 | 291.73 | 1046.71 | 1050.92 | 657.48 | 0.996 | 1.592 | 0.996 | 0.626 |
| 38 | 334.18 | 1338.04 | 1328.74 | 670.02 | 1.007 | 1.997 | 1.007 | 0.504 |
| 40 | 337.37 | 1102.86 | 1101.76 | 869.77 | 1.001 | 1.268 | 1.001 | 0.789 |

For an isotropic solid, the three-dimensional (3D) direction-dependent Young's modulus, linear compressibility, shear modulus, and Poisson's ratio should all have spherical shapes. on the other hand, anisotropy is indicated by any divergence from the sphere. **Figures 6** to **9** depict the two-dimensional and three-dimensional variations in elastic moduli (*Y*), shear modulus (*G*), linear compressibility (*β*), and Poisson's ratio (*v*), created using the ELATE tool [92]. This matrix is placed into the ELATE software to visualize the *Y*, *G*, *v*, and *β* anisotropy levels of La$_3$Ni$_2$O$_7$ material. The green and blue shapes correspond to the minimum and maximum values of that polycrystalline property, respectively, while the red shapes represent negative values. The maximum and minimum limits for these variables are presented in **Table 8**. As can be seen from the figures there is a small deviation from spherical shape in the 3D figures of $Y$, $G$, $\beta$, and $v$ for all the pressures signifying some degree of anisotropy.

**Table 8.** The lower and upper bounds of Young's modulus ($Y$ in GPa), linear compressibility ($\beta$ in TPa$^{-1}$), shear modulus ($G$ in GPa), and Poisson's ratio ($v$) for the La$_3$Ni$_2$O$_7$ compound.

| Pressure (GPa) | Y | | $A_Y$ | $\beta$ | | $A_\beta$ | G | | $A_G$ | v | | $A_v$ |
|---|---|---|---|---|---|---|---|---|---|---|---|---|
| | $Y_{min}$ | $Y_{max}$ | | $\beta_{min}$ | $\beta_{max}$ | | $G_{min}$ | $G_{max}$ | | $v_{min}$ | $v_{max}$ | |
| 30 | 254.93 | 372.83 | 1.46 | 0.93 | 1.29 | 1.39 | 92.16 | 129.09 | 1.40 | 0.24 | 0.44 | 1.80 |
| 32 | 258.31 | 393.36 | 1.52 | 1.07 | 1.25 | 1.17 | 93.63 | 170.53 | 1.82 | 0.15 | 0.48 | 3.14 |
| 34 | 269.51 | 405.73 | 1.51 | 1.01 | 1.22 | 1.22 | 98.60 | 174.86 | 1.77 | 0.16 | 0.48 | 2.97 |
| 36 | 202.50 | 304.55 | 1.50 | 0.95 | 1.52 | 1.60 | 83.22 | 128.05 | 1.54 | 0.19 | 0.52 | 2.75 |
| 38 | 225.77 | 342.43 | 1.52 | 0.75 | 1.49 | 2.00 | 92.60 | 138.04 | 1.49 | 0.24 | 0.50 | 2.09 |
| 40 | 299.47 | 448.46 | 1.50 | 0.91 | 1.15 | 1.27 | 110.09 | 190.39 | 1.73 | 0.18 | 0.46 | 2.61 |

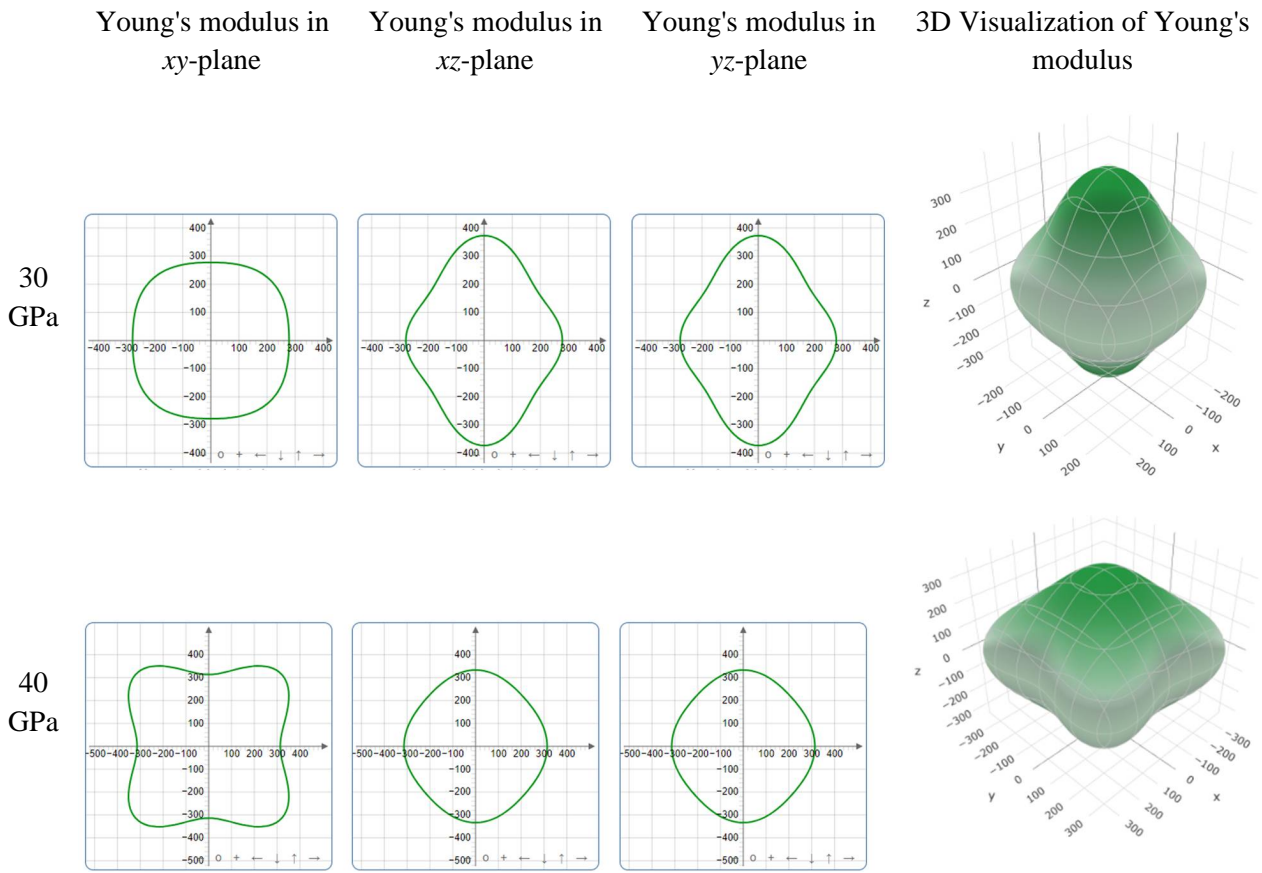

**Figure 6.** Directional variation in Young's modulus ($Y$) of La$_3$Ni$_2$O$_7$ compound.

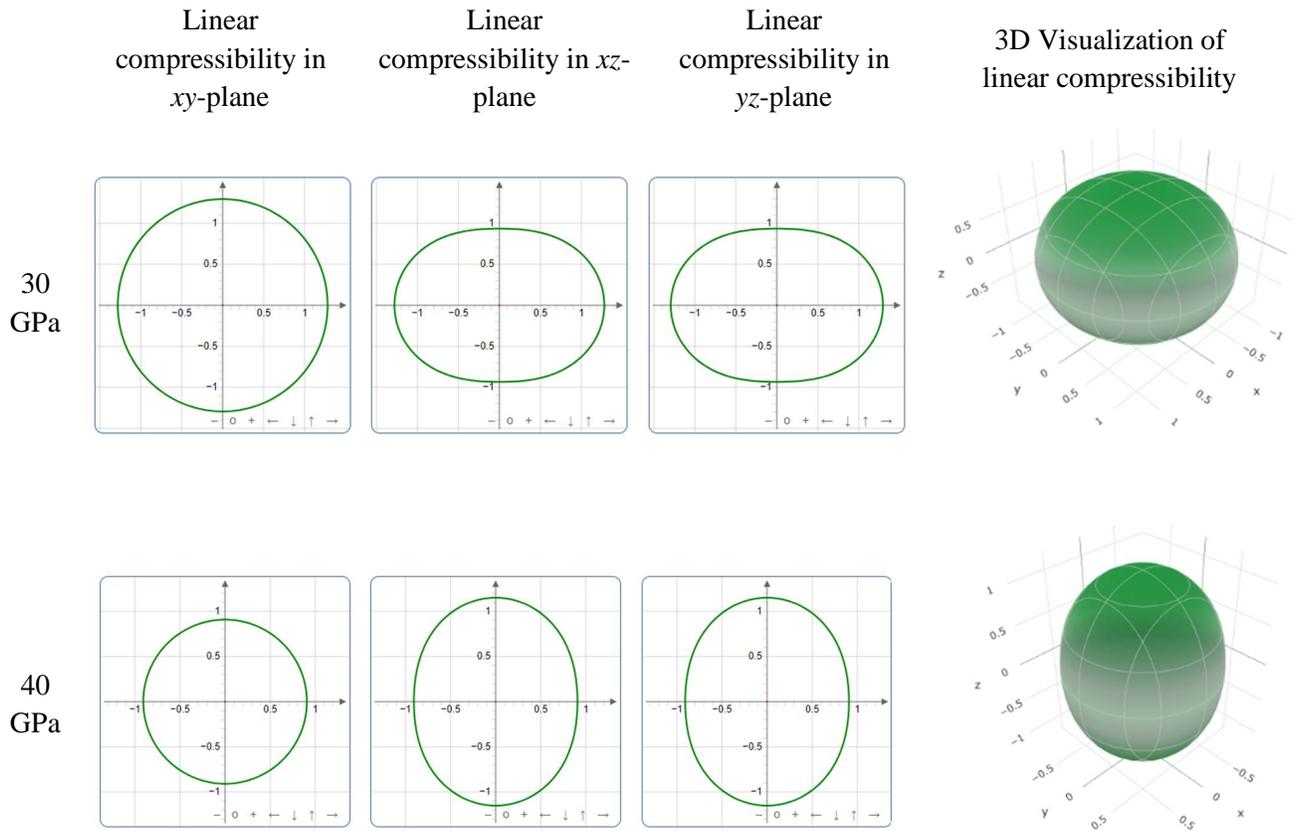

**Figure 7.** Directional variation in linear compressibility (β) of La$_3$Ni$_2$O$_7$ compound.

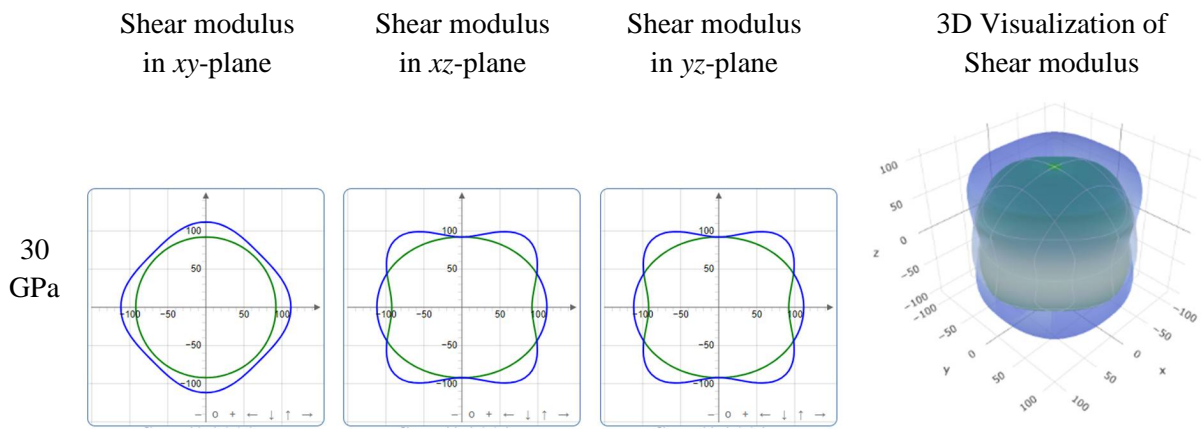

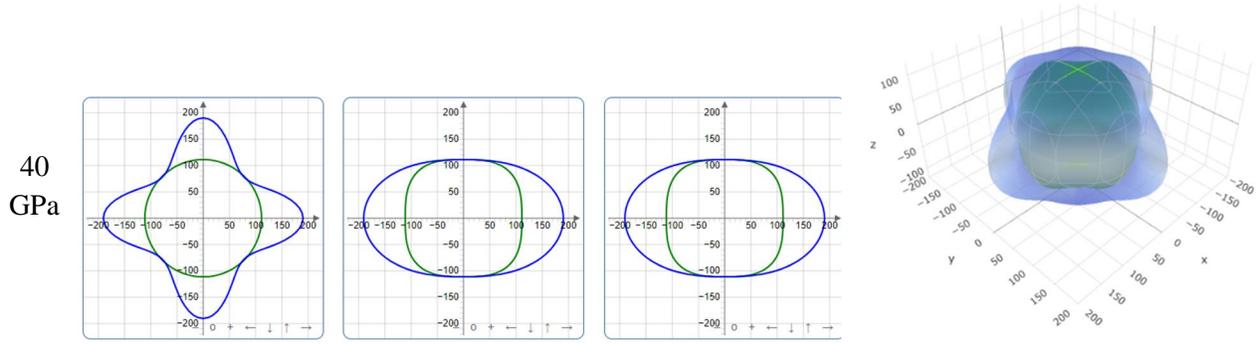

**Figure 8.** Directional variation in shear modulus ($G$) of La$_3$Ni$_2$O$_7$ compound.

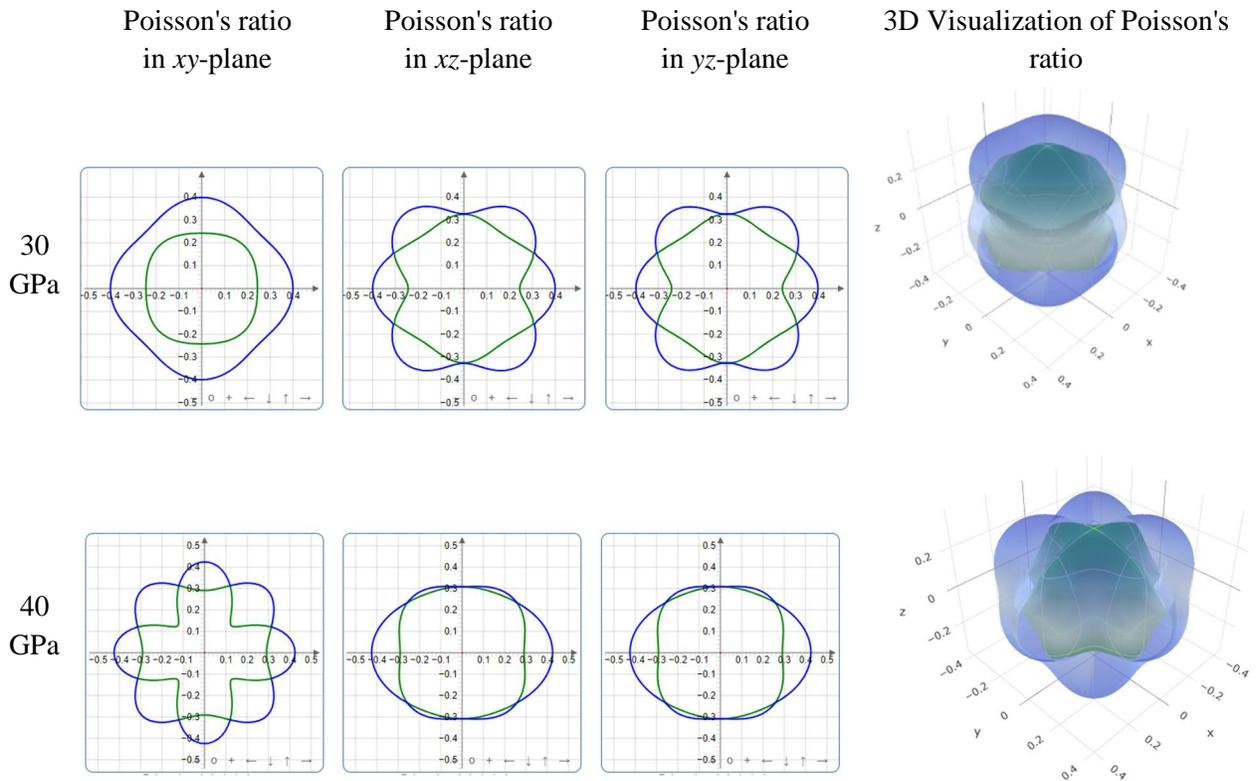

**Figure 9.** Directional variation in Poisson's ratio ($v$) of La$_3$Ni$_2$O$_7$ compound.

### 3.4 Electronic Properties

**(a) Band Structure**

The valence and conduction electrons in a material determine nearly all of its physical properties that are relevant to technology and fundamental physics. The behavior of these electrons is determined by the particular form of their energy dispersion, ($E(k)$), throughout the Brillouin zone. The changes in electronic energy across different bands as a function of momentum outline the electronic band structure. The electronic band structure is a crucial concept in solid-state physics, helping to explain various material properties, including electrical conductivity, electronic thermal conductivity, electronic heat capacity, the Hall effect, magnetic properties, and optoelectronic characteristics. Using the optimized crystal structure of La$_3$Ni$_2$O$_7$, we have determined its electronic band structure. **Figure 10** (a-f) illustrates the electronic energy dispersion curves at various pressures along the high-symmetry directions of the Brillouin zone for La$_3$Ni$_2$O$_7$. The horizontal broken red line marks the Fermi level ($E_F$).

The band structure clearly shows that several bands, each with different levels of dispersion, pass through the Fermi level. This affirms the metallic nature of the La$_3$Ni$_2$O$_7$ compound. This result is consistent with the earlier findings [18,51,93]. Bands crossing the Fermi level are indicated in colored lines with their corresponding band numbers. The band numbers that cross the Fermi level are 189, 190, 191, and 192. Highly dispersive bands running along Z-T, Y-S, and, S-X directions show electronic character for all pressures. These bands are expected to control the properties of charge transport. The bands that intersect the Fermi level near the G-Z and mid of Z-T symmetry points are significantly less dispersive almost like flat bands for all pressures. These flat band electrons are strongly correlated. Bands that disperse strongly suggest a low effective mass of charge carriers and high mobility of charges [94–96]. The material being studied demonstrates significant anisotropy in charge transport along different crystallographic directions and in momentum space.

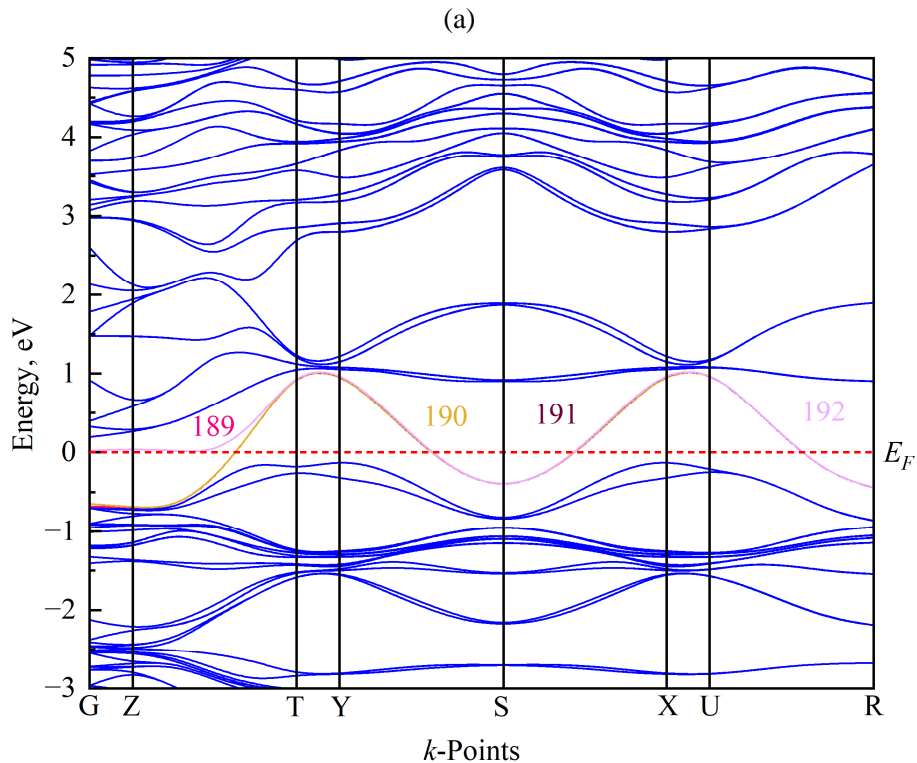

(a)

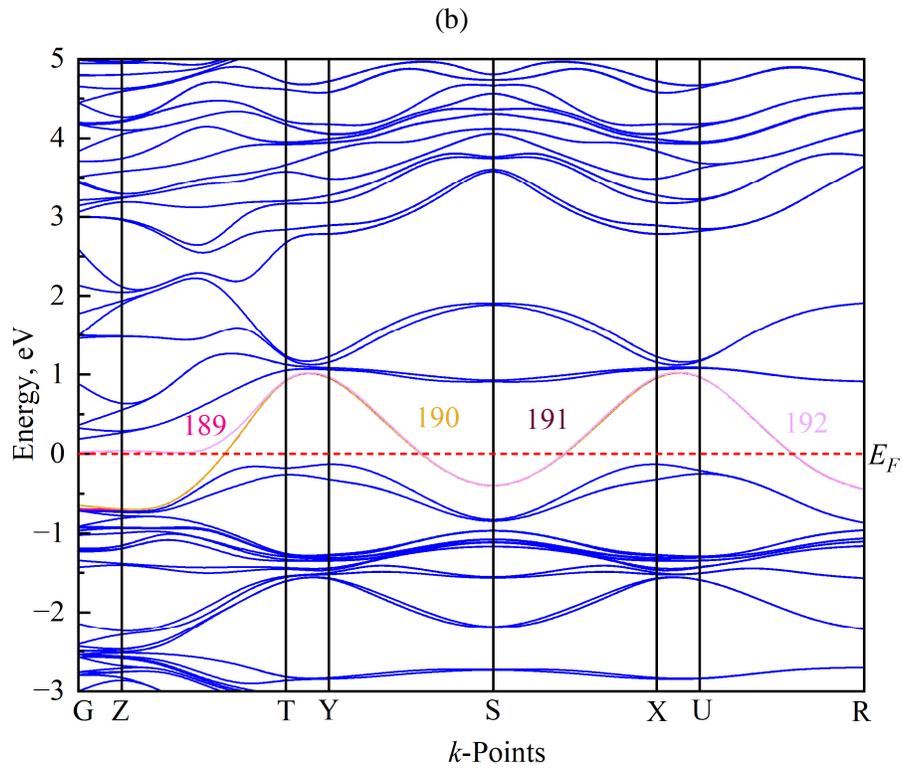

(b)

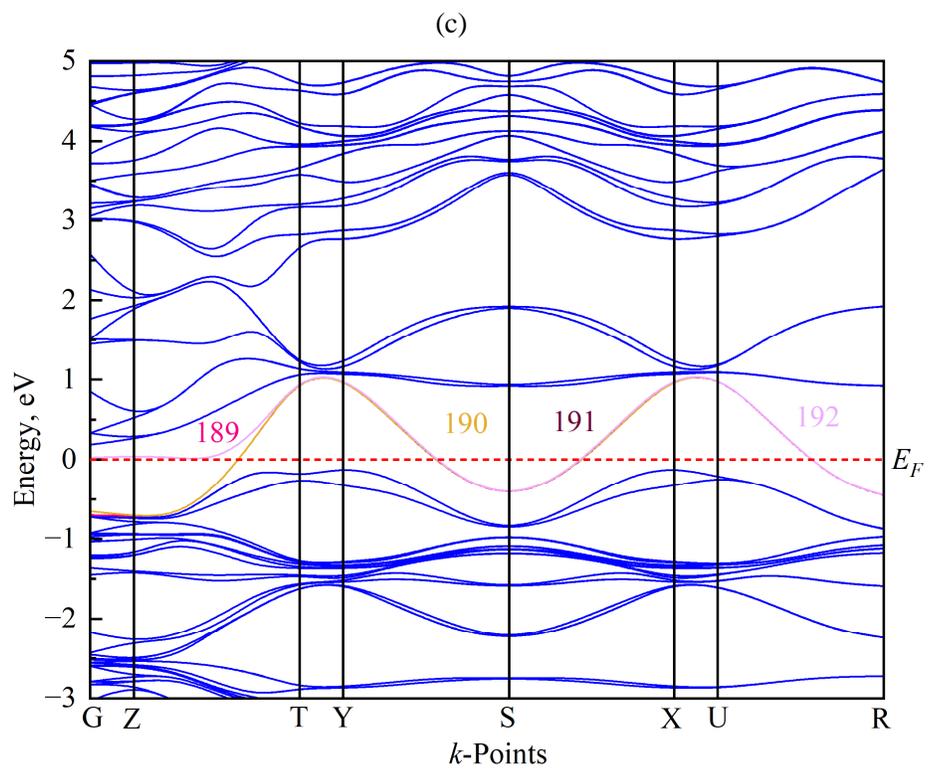

(c)

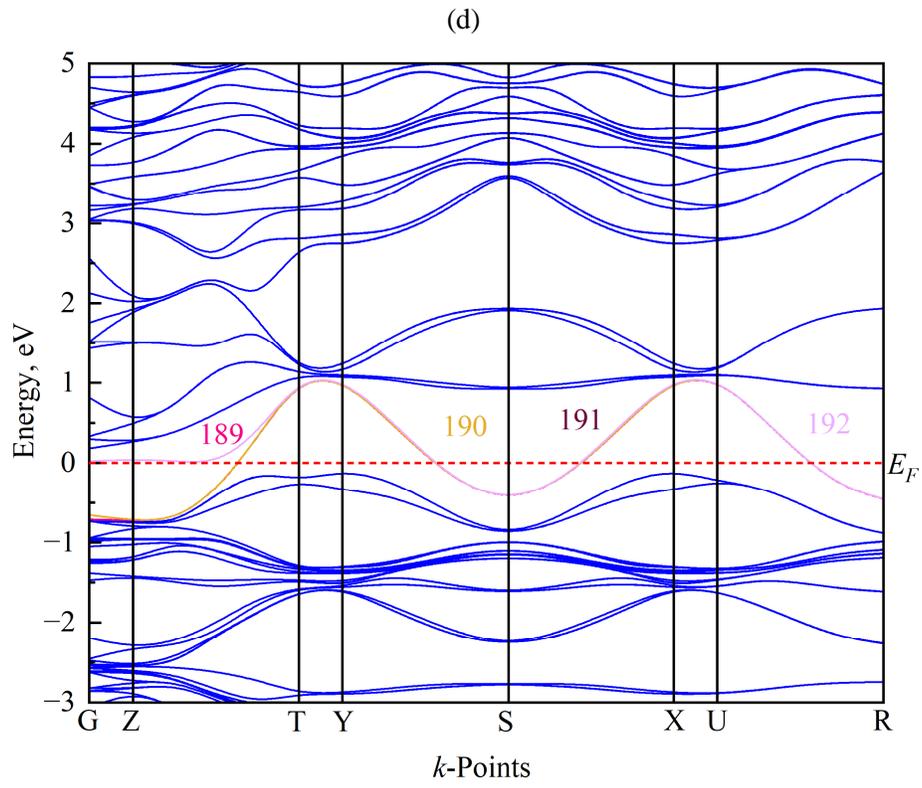

(d)

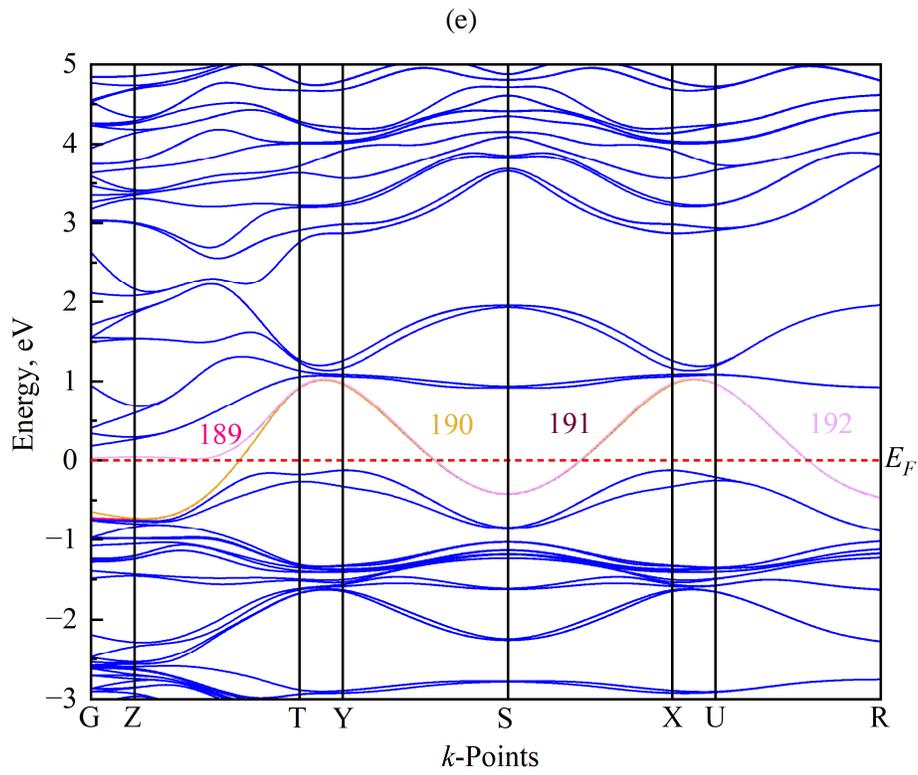

(e)

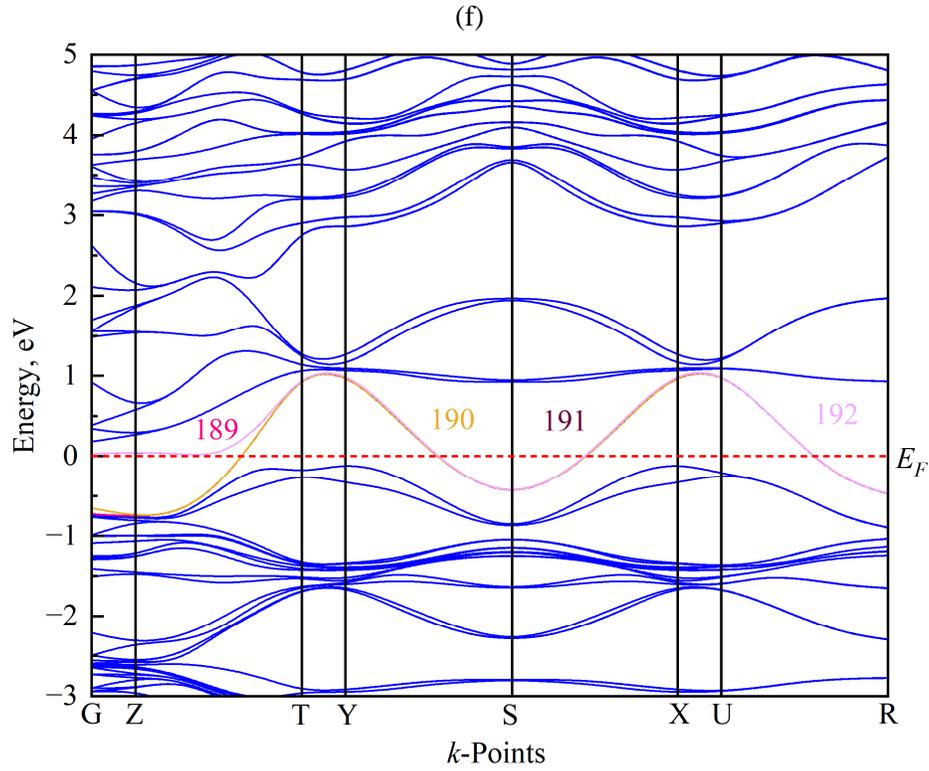

**Figure 10.** The electronic band structure of La$_3$Ni$_2$O$_7$ along the high symmetry directions of the *k*-space within the first Brillouin zone at pressures of (a) 30 GPa, (b) 32 GPa, (c) 34 GPa, (d) 36 GPa (e) 38 GPa (f) 40 GPa.

**Figure 11** presents the band structure of the La$_3$Ni$_2$O$_7$ compound at different pressures, providing a comparison of how the band structure changes with pressure. The electronic band structure generally changes noticeably in response to large changes in the cell volume. This is because the periodicity of the ionic potential alters, impacting both the nature and extent of band dispersion. Unexpectedly, there is minimal variation in the ($E(k)$) features with changes in pressure. (**Figure 11**). One possible explanation is that the pressure coefficient of the deformation potential has opposite signs along different crystallographic axes, and when combined, they tend to cancel each other out, leading to an almost negligible overall impact of pressure on the electronic band structure.

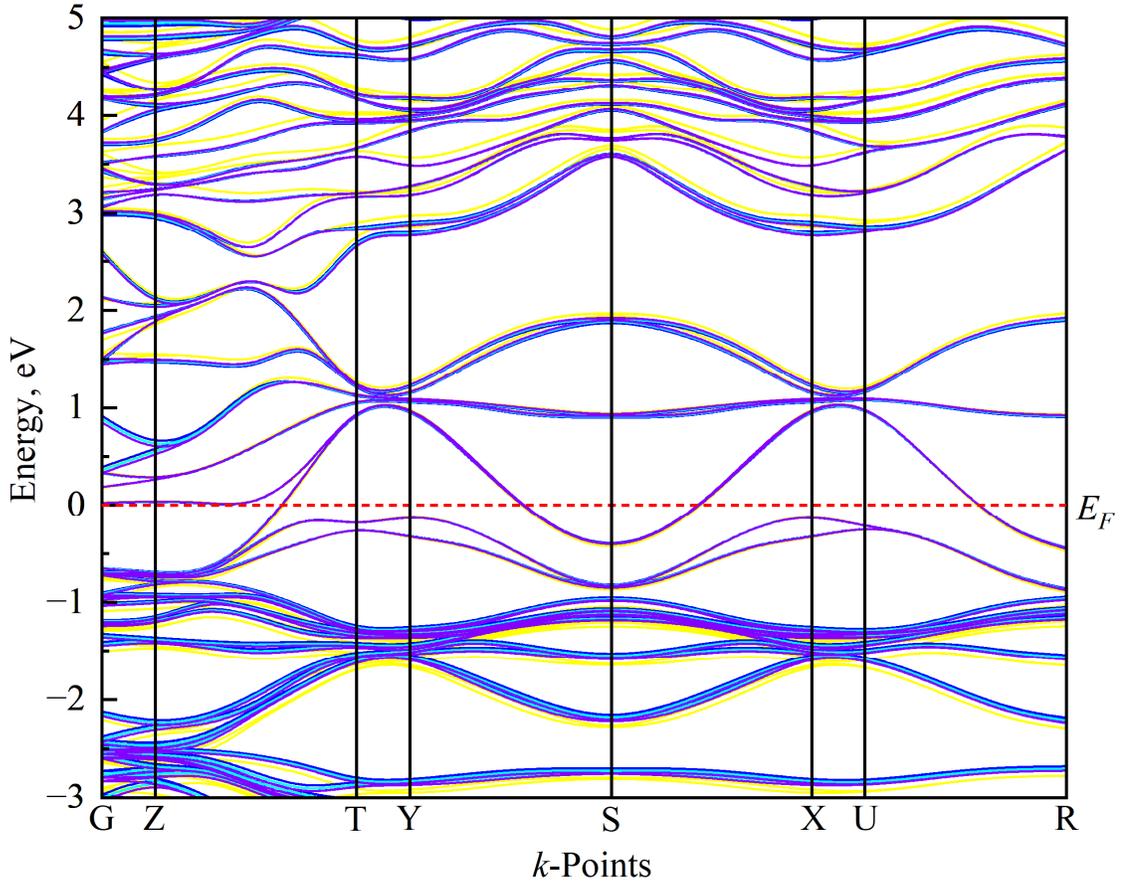

**Figure 11.** Comparison of band structures of La$_3$Ni$_2$O$_7$ along high symmetry directions in the first Brillouin zone (blue, cyan, violet, dark yellow, purple, and yellow colors represent energy dispersion curves for 30, 32, 34, 36, 38, and 40 GPa pressures, respectively).

**(b) Electronic Energy Density of States (EDOS)**

The electronic density of states (DOS) of a system represents the number of electronic states available to be occupied at each energy level per unit energy interval. Nearly every electronic and optical characteristic of a crystalline solid is dependent on the DOS's structure in the valence and conduction bands. Furthermore, it's critical to comprehend a material's DOS in order to comprehend how each atom contributes to bonding and antibonding states [61]. It should be noted that the group velocity of the charge carriers is determined by the derivative of the energy dispersion curves (band structures) with respect to momentum ($k$). The density of states and this first derivative are also connected; where the derivative is low, the DOS is high, and vice versa [43]. The effective mass of electrons/holes is directly proportional to the inverse of the second derivative of the $E(k)$ curves. The nature of curvature of the $E(k)$ curves characterizes the electron- and hole-like bands. These characteristics have been employed to relate the band structure to the DOS and to interpret the fundamental aspects of energy dispersion curves. In **Figure 12**, the total and partial density of states (TDOS and PDOS) are shown as functions of energy ($E-E_F$), obtained from the band dispersions under varying pressures. The vertical dashed blue line represents the Fermi energy, ($E_F$) which is set to zero. The compound under investigation should demonstrate metallic electrical conductivity, as indicated by the non-zero value of TDOS at the Fermi level. This outcome is consistent with the earlier findings [18,51,93]. To better understand the contribution of each electronic orbital to the TDOS and their role in atomic bonding, we have computed the orbital-resolved PDOS for La, Ni, and O atoms in $La_3Ni_2O_7$ under different pressures, as shown in **Figure 12** (a-f). The values of TDOS at the Fermi level are 11.94, 11.85, 11.87 11.79, 11.89, 11.93 states/eV- unit cell at 30 GPa, 32 GPa, 34GPa, 36 GPa, 38 GPa, and 40 GPa, respectively. Finite TDOS at the Fermi level again confirms the metallicity of $La_3Ni_2O_7$. The majority of the TDOS, or roughly 67 %, is contributed by the Ni 4*d*. This suggests that the characteristics of the Ni 3*d* electronic states primarily affect the chemical and mechanical stability of $La_3Ni_2O_7$. The primary contribution to the TDOS near $E_F$ comes from Ni 3*d* and O 2*p* states in which O 2*p* contributes to the TDOS roughly about 23%, with a small additional contribution from the La 6*s* and La 5*d* electronic states. Therefore, there is considerable hybridization between the Ni 3*d* electronic states and the O 2*p* electronic states close to the Fermi energy. This is similar to the case of hole doped cuprates. Strong hybridization and increased DOS suggest that Ni and O atoms may form covalent bonds. In comparison to the PDOS from Ni and O, the PDOS from La is significantly smaller, both in the valence band below the Fermi level and the conduction band above it. The small PDOS and weaker hybridization between La 6*s*, La 5*d*, and the atomic orbitals of Ni and O indicate that the bonding between these atoms is not strong. There are a number of peaks in the TDOS, e.g., at ~ -1.12 eV, 1.11 eV, 1.93 eV, and 3.46 eV. Each of these peaks is expected to play a key role in optical transitions in the visible region. In particular, the peaks near the Fermi level are expected to govern charge transport and the associated electrical properties. The large peak at approximately -1.12 eV being near the Fermi energy suggests that $La_3Ni_2O_7$ could be a promising material for band engineering.

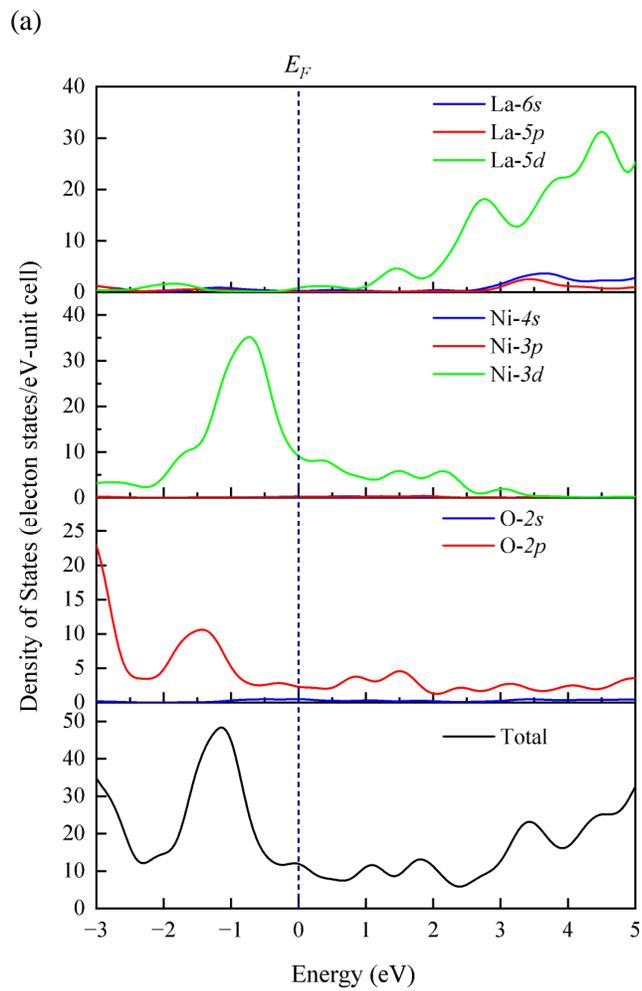
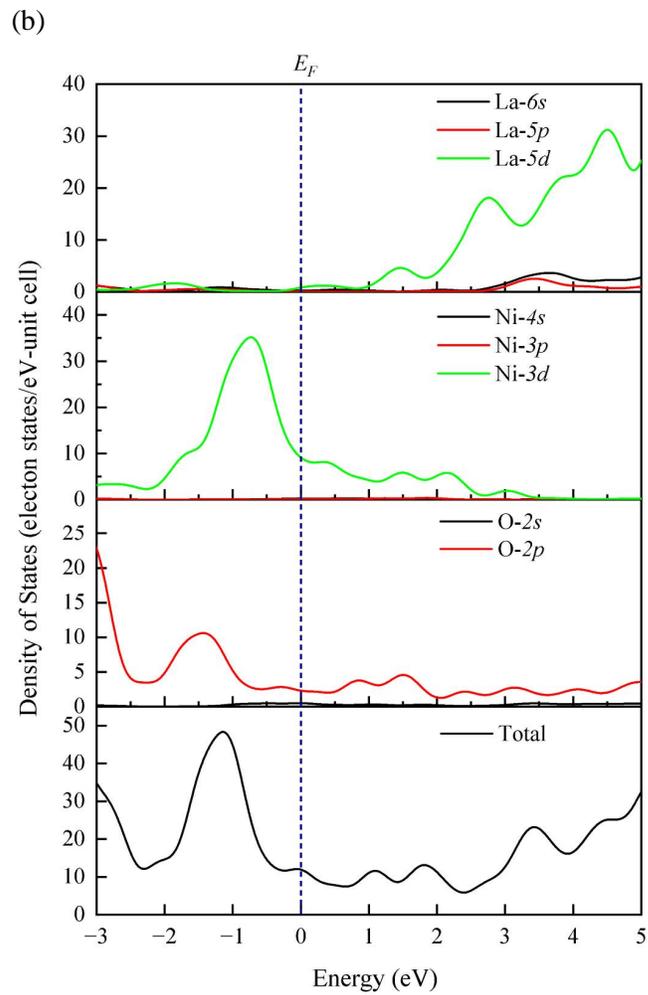

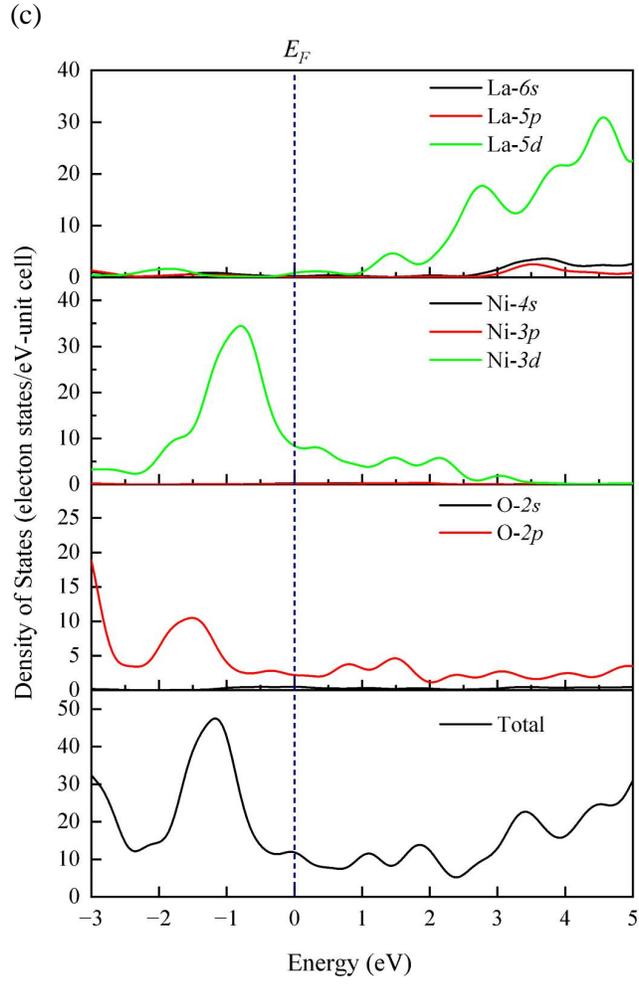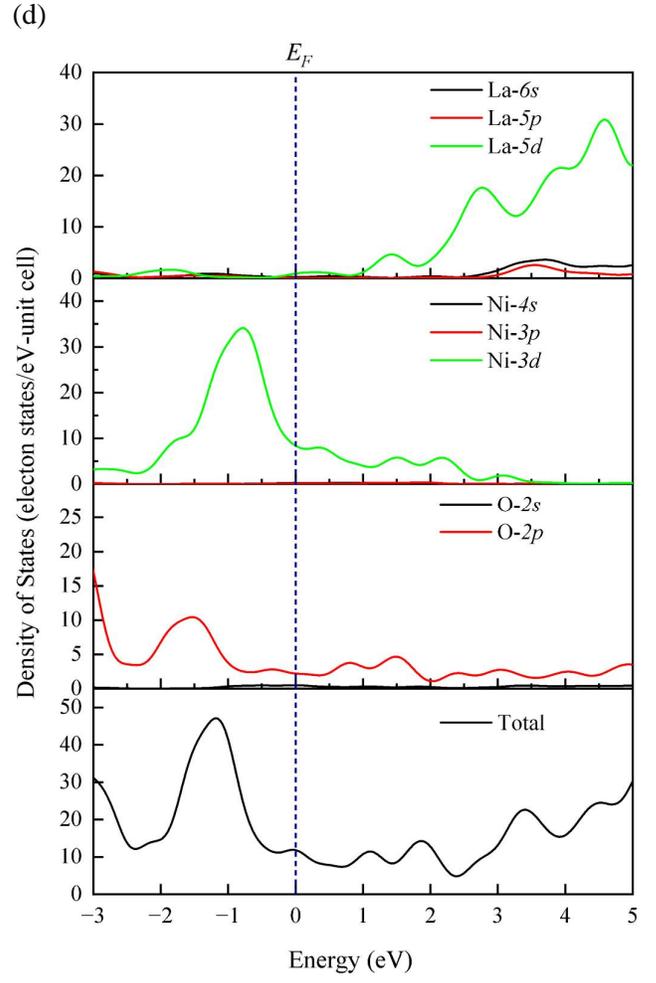

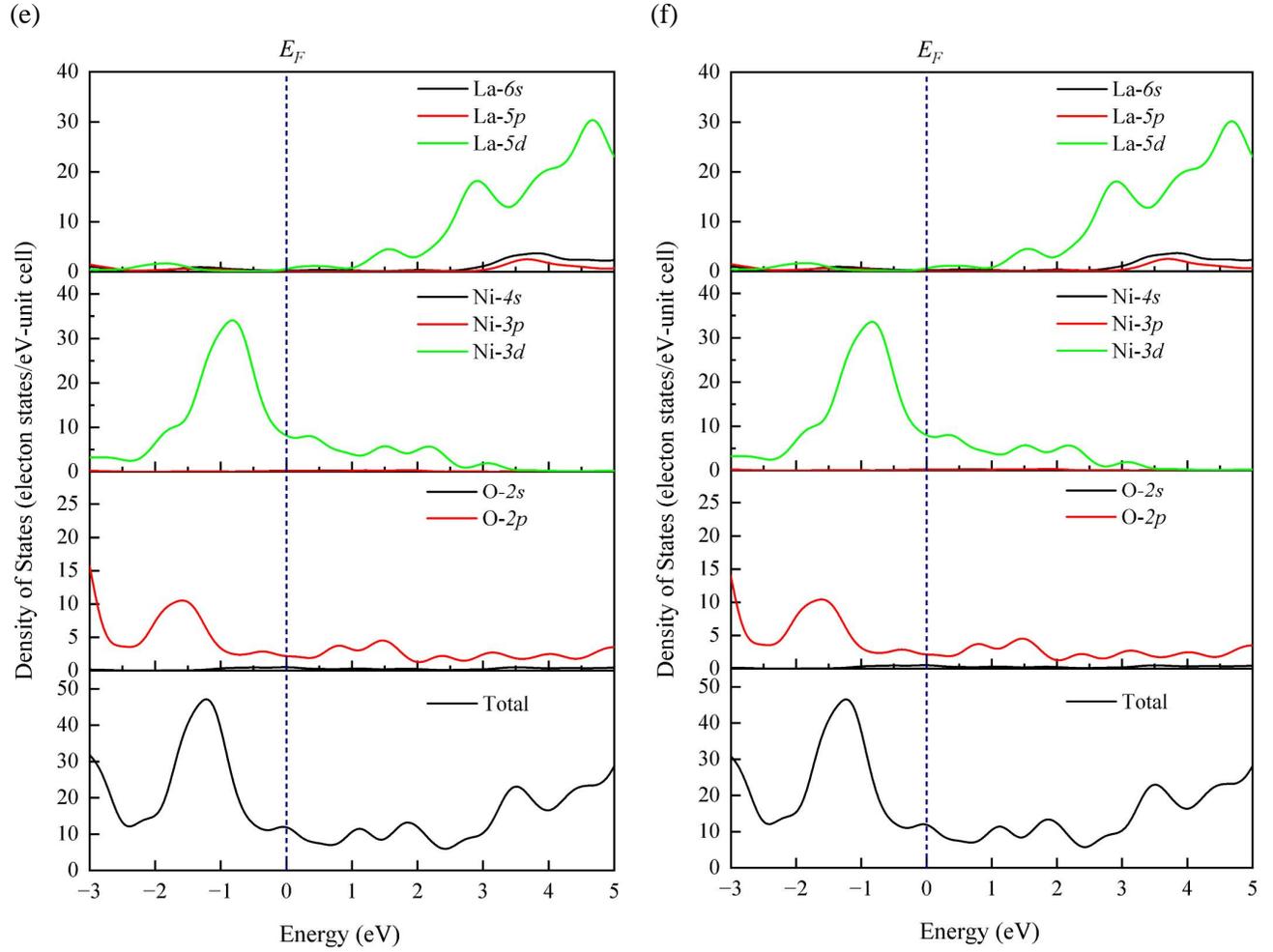

**Figure 12.** The total and partial electronic density of states of La$_3$Ni$_2$O$_7$ at various pressures: (a) 30 GPa, (b) 32 GPa, (c) 34 GPa, (d) 36 GPa, (e) 38 GPa, and (f) 40 GPa.

In **Figure 13**, the total density of states of the La$_3$Ni$_2$O$_7$ compound at different pressures is presented, enabling a comparison of the variations induced by pressure. Altogether it can be asserted that applied pressure appreciably alters neither the band structure nor the TDOS.

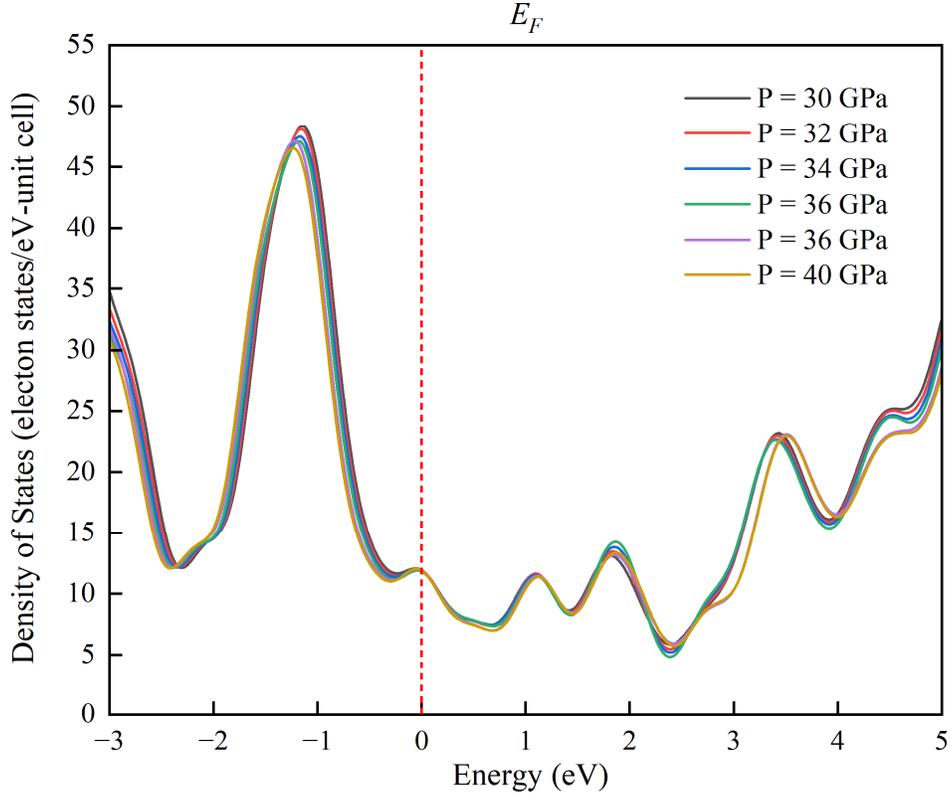

**Figure 13.** Comparison of the total density of states of La$_3$Ni$_2$O$_7$ at pressures of (a) 30 GPa, (b) 32 GPa, (c) 34 GPa, (d) 36 GPa, (e) 38 GPa, and (f) 40 GPa.

**(c) Coulomb Pseudopotential**

The electron-electron interaction parameter, commonly known as the Coulomb pseudopotential parameter, for a compound can be roughly approximated using the following equation [97,98]:

$$\mu^* = \frac{0.26 N(E_F)}{1 + N(E_F)} \tag{32}$$

This parameter measures the repulsive Coulomb correlations in solids. The calculated values of $\mu^*$ for La$_3$Ni$_2$O$_7$ at various pressures are shown in **Table 9** below.

**Table 9.** Calculated values of the Coulomb pseudopotential, $\mu^*$, and DOS at the Fermi level, $N(E_F)$ (electron states/eV-unit cell) of La$_3$Ni$_2$O$_7$ for different values of pressure.

| Pressure (GPa) | DOS at the Fermi level | Coulomb pseudopotential |
|---|---|---|
| 30 | 11.939 | 0.195 |

| 32 | 11.846 | 0.194 |
| 34 | 11.870 | 0.194 |
| 36 | 11.793 | 0.194 |
| 38 | 11.888 | 0.195 |
| 40 | 11.927 | 0.195 |

In comparison to many conventional superconductors, which typically have values between 0.10 and 0.15 [99], this value is quite large for our investigated pressure range. It is important to observe that the Coulomb pseudopotential remains approximately unchanged as pressure increases. In the context of superconductivity, the repulsive Coulomb pseudopotential diminishes the effective electron-phonon interaction that drives the formation of Cooper pairs [99,100]. Consequently, the superconducting transition temperature, $T_c$, is suppressed [97,99,101].

**(e) Fermi Surface**

The topology of the Fermi surface (FS) governs the electronic properties of metals. The Fermi surface separates the occupied and unoccupied electronic states at low temperatures. Electronic, optical, thermal, and magnetic properties are among the many that are greatly influenced by the topology of a Fermi surface. The Fermi surface topology of $La_3Ni_2O_7$ is shown in **Figure 14 - 15** (a, b, c, d, and e). The bands 189, 190, 191, and 192 cross the Fermi level and are responsible for the formation of the Fermi surface (**Figure 10** (a-f)). The Fermi surface of $La_3Ni_2O_7$ is made up of four Fermi sheets of different shapes. The FSs for the bands 189 and 190 are quite similar. Both structures have electron-like sheets that appear along the T-Y and X-U paths for both pressures. On the other hand, the FSs for 191 and 192 bands are also similar for pressure 30 GPa and comparatively complex in shape. But surprisingly the FSs for 191 and 192 bands for pressure 40 GPa are slightly dissimilar from one another where there is a hole in the Fermi sheet for band 192. The Fermi sheet enclosing the central sheet has a hole-like character. The central sheet is electronic and appears along the G-Z path. This implies that both electron- and hole-like behaviors exist in $La_3Ni_2O_7$. The Fermi surface topology also indicates that electronic transport should be anisotropic in $La_3Ni_2O_7$.

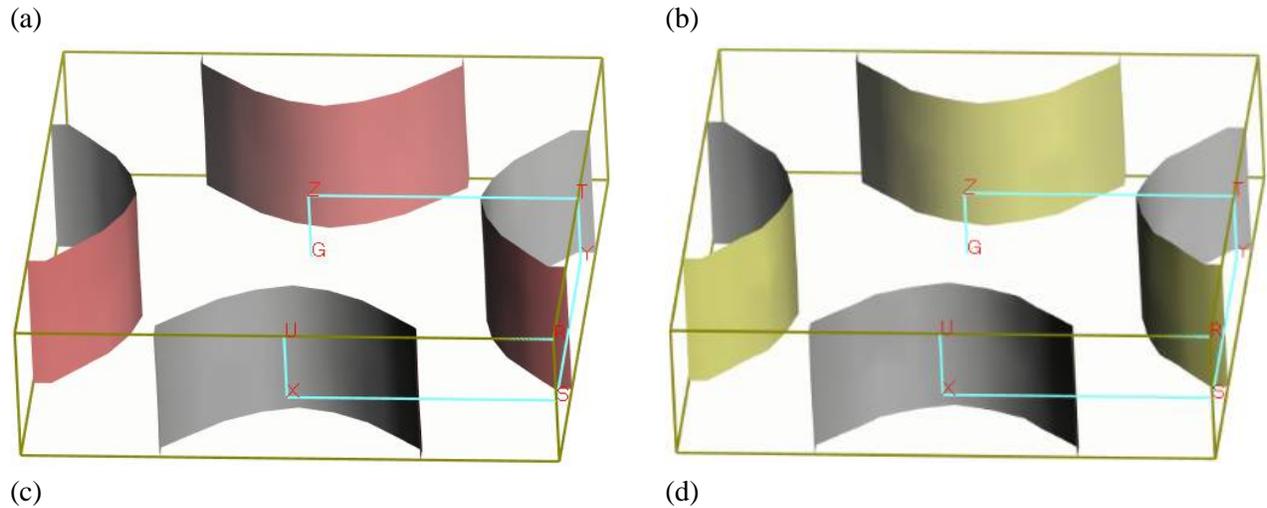

(a)                          (b)

(c)                          (d)

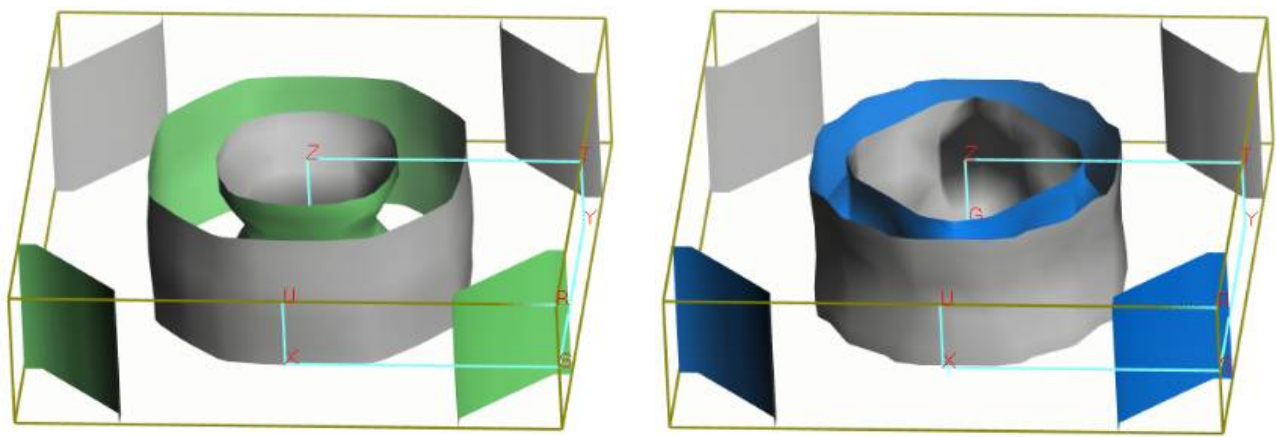

(e)

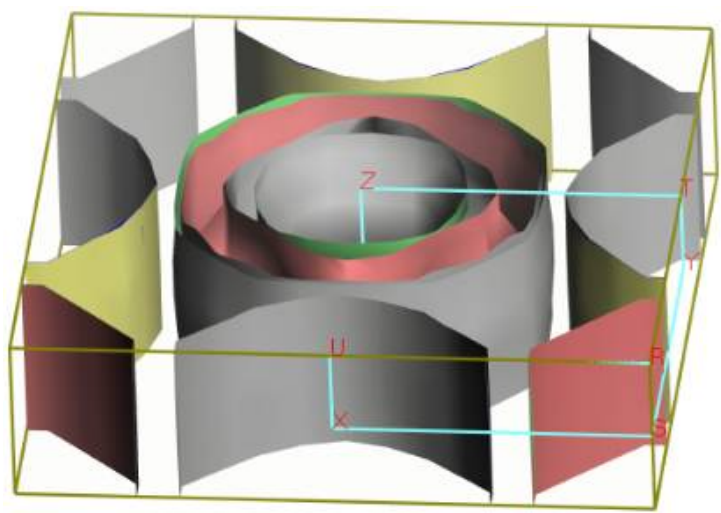

**Figure 14.** Fermi surface for bands (a) 189 (b) 190 (c) 191 (d) 192 and (e) of La$_3$Ni$_2$O$_7$ compound at pressure 30 GPa.

(a) (b)

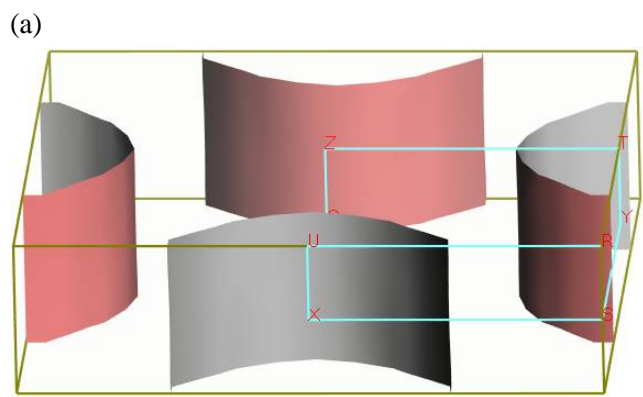
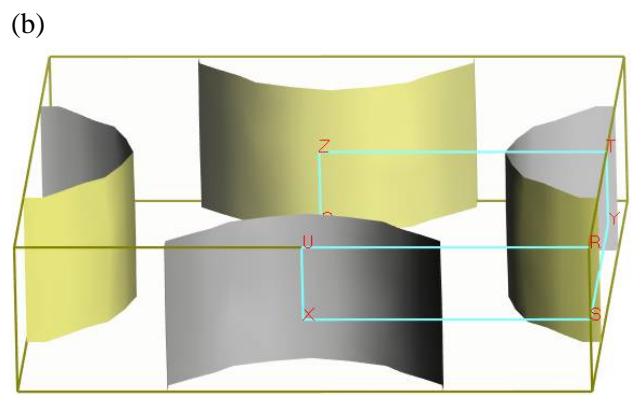

(c) (d)

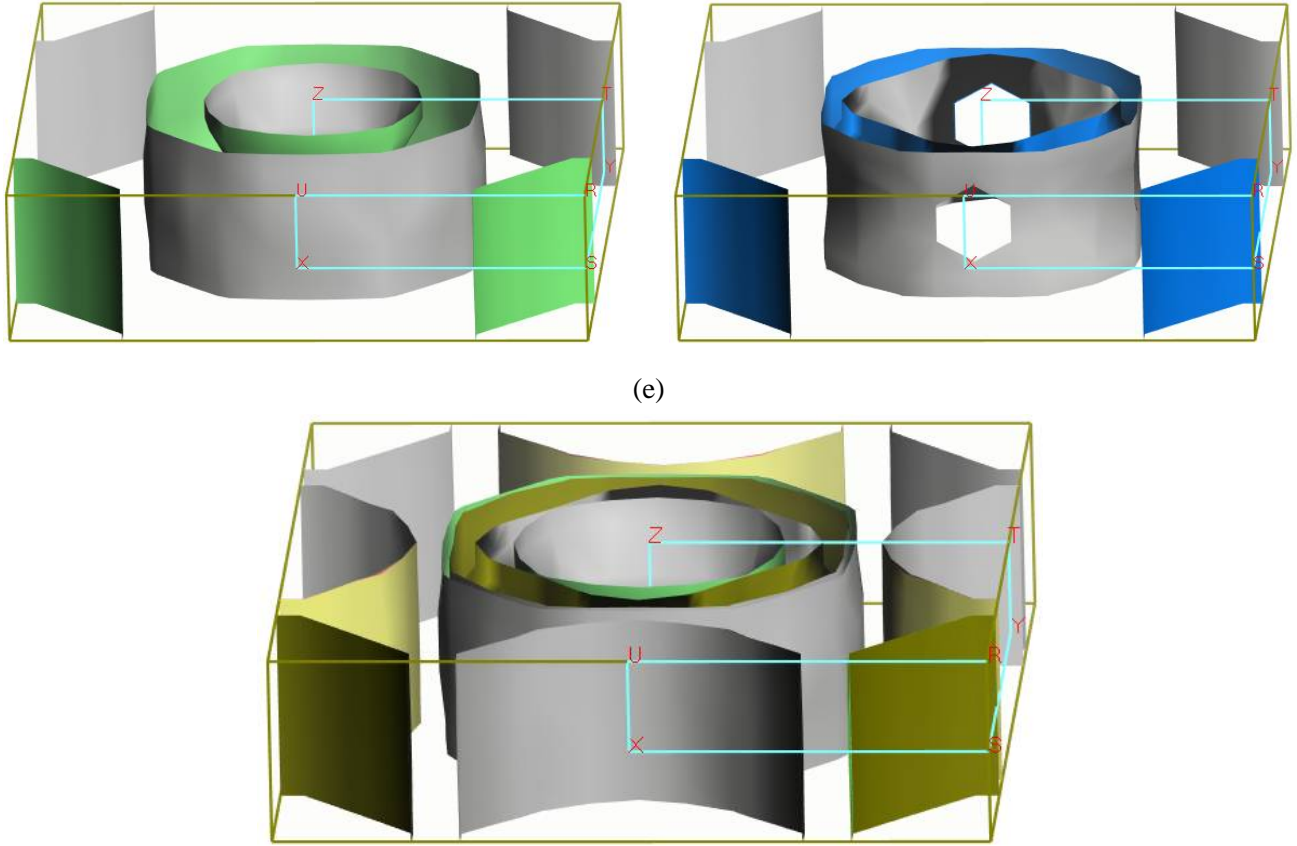

**Figure 15.** Fermi surface for bands (a) 189 (b) 190 (c) 191 (d) 192 and (e) of La$_3$Ni$_2$O$_7$ compound at pressure 40 GPa.

### 3.5 Thermophysical Properties

The analysis of thermal properties such as Debye temperature, melting temperature, lattice thermal conductivity, minimum thermal conductivity, and thermal expansion coefficient is vital for predicting the thermodynamic stability and evaluating its potential for various applications.

**(a) Debye temperature**

The Debye temperature, $\theta_D$, is a key thermo-physical parameter of solids. The temperature at which all of the atomic modes of vibration become active is known as the Debye temperature. The Debye frequency, $v_D$, is the frequency corresponding to the highest energy vibrational mode. The relationship between $\theta_D$ and $v_D$ is given by the equation $hv_D = k_B \theta_D$, where $h$ is the Planck constant and $k_B$ is the Boltzmann constant. The Debye temperature is influenced by various physical parameters, with the average inter-atomic force being a key factor. In general, a stiffer structure leads to a higher Debye temperature. In addition, $\theta_D$ is closely associated with various properties, including melting temperature, phonon-specific heat, phonon thermal conductivity, superconducting transition temperature in phonon-mediated superconductors, and vacancy formation energy in crystals, among others [102,103]. There are various methods for determining the value of $\theta_D$. In this section, we use the approach developed by Anderson.

Using the elastic moduli of solids, this technique produces accurate estimates of the Debye temperature and is widely applied to calculate $\theta_D$ for solids with varying electronic band structures and crystal symmetries [104,105]. According to Anderson the expression for $\theta_D$ is given by [106]:

$$\theta_D = \frac{h}{k_B}\left[\left(\frac{3n}{4\pi}\right)\frac{N_A\rho}{M}\right]^{\frac{1}{3}} v_m \qquad (33)$$

Here, $h$ is Planck's constant, $k_B$ is the Boltzmann constant, $N_A$ is Avogadro's number, $\rho$ is the density, $M$ is the molecular weight, $n$ is the number of atoms in the molecule, and $v_m$ is the mean sound velocity in a given medium. The average sound velocity $v_m$ in the solid is determined from,

$$v_m = \left[\frac{1}{3}\left(\frac{1}{v_l^3} + \frac{2}{v_t^3}\right)\right]^{-\frac{1}{3}} \qquad (34)$$

where $v_l$ and $v_t$ represent the longitudinal and transverse sound velocities, respectively. The longitudinal and transverse sound velocities can be derived from the bulk modulus $B$ and shear modulus $G$ using the following equations:

$$v_l = \left[(B + \frac{4G}{3})/\rho\right]^{\frac{1}{2}} \qquad (35)$$

and,

$$v_t = \left[\frac{G}{\rho}\right]^{\frac{1}{2}} \qquad (36)$$

The calculated values of sound velocities and $\theta_D$ at different pressures are presented in **Table 10** and also shown in **Figure 16**. Generally, the sound velocities rise with increasing pressure because of the enhanced crystal stiffness and density [107]. Once again, we observe an unusual decrease in sound velocities at a specific pressure. For each direction of propagation, the longitudinal sound velocities exceed the transverse velocities. This occurs because $C_{11}$ is greater than both $C_{12}$ and $C_{44}$ in La$_3$Ni$_2$O$_7$. From **Table 10** [cf. **Figure 16** (a)], it can be inferred that the $\theta_D$ increases monotonously with pressure again except at certain pressure. In general, crystal stiffening is indicated by an increase in $\theta_D$ with pressure; however, in the opposite case, the system tends to experience lattice softening [108]. The anomalous variation of $G$ with pressure may account for the unexpected change in $\theta_D$ (cf. **Table 4** and **Table 10**). It is worth emphasizing that the high predicted $T_c$ of La$_3$Ni$_2$O$_7$ can be partially explained by its high Debye temperature. The superconducting critical temperature in conventional phonon-mediated superconductors is directly related to the Debye temperature. The pressure-induced rise in the Debye temperature observed for La$_3$Ni$_2$O$_7$ indicates that $T_c$ should increase with pressure, as long as the electron-phonon coupling constant does not significantly decrease with pressure changes.

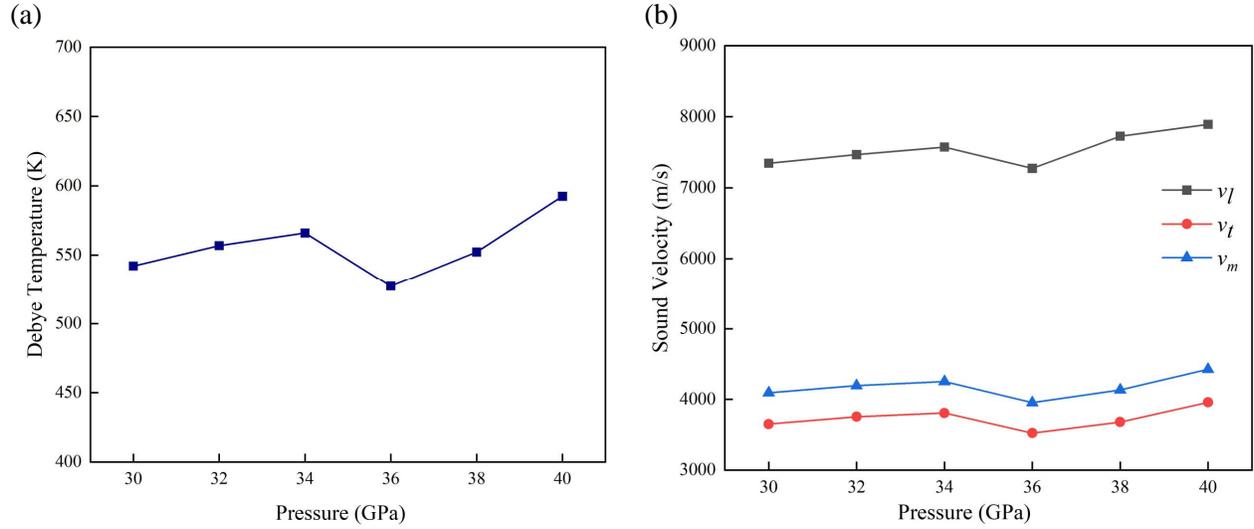

**Figure 16.** (a) Debye temperature and (b) Sound velocities of $La_3Ni_2O_7$ under pressure.

**(b) Heat Capacity**

The heat capacity of a material is a crucial intrinsic thermodynamic property. A material with a high heat capacity will have a low thermal diffusivity and a high thermal conductivity. The heat capacity per unit volume ($\rho C_P$) represents the amount of thermal energy change per unit volume of a material for a one-degree Kelvin change in temperature. The heat capacity per unit volume ($\rho C_P$) of a material is given by [85,109]:

$$\rho C_P = \frac{3k_B}{\Omega} \tag{37}$$

where $N = 1/\Omega$ denotes the number of atoms per unit volume. Table 10 presents the heat capacity per unit volume of $La_3Ni_2O_7$. It is evident from **Table 10** [cf. **Figure 17** (b)] that heat capacity per unit volume increases monotonically with increased pressure.

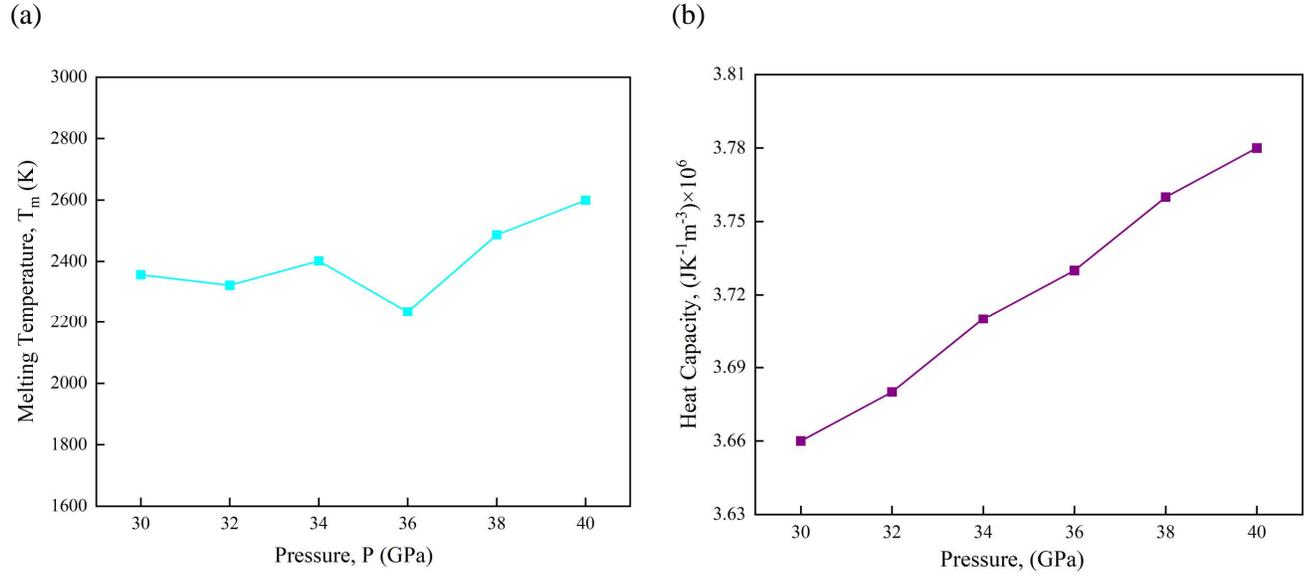

**Figure 17.** (a) Variation of (a) melting temperature and (b) heat capacity of $La_3Ni_2O_7$ at different pressures.

**(c) Melting Temperature**

The melting temperature ($T_m$) is an essential thermophysical parameter that indicates the possibility of a material's application at higher temperatures. High cohesive energy, bonding energy, a low coefficient of thermal expansion, and strong atomic interaction are all present in crystals with high melting temperatures [91]. Solids can be used continuously at temperatures below $T_m$ without oxidation, chemical change, or excessive distortion that could lead to mechanical or elastic failure. The melting temperature $T_m$ of the $La_3Ni_2O_7$ structure is estimated based on its elastic constants using the following expression. [66,109]:

$$T_m = [354 + 1.5(2C_{11} + C_{33})] \, K \qquad (38)$$

The calculated values of $T_m$ for $La_3Ni_2O_7$ are disclosed in **Table 10** [cf. **Figure 17** (a)]. It is found from **Figure 17** (a) that the melting point rises with the increase of pressure though there is a slight fall at certain pressures. The melting temperatures of $La_3Ni_2O_7$ are quite high for all pressures. The bonding strength of a crystal is closely linked to its $T_m$. The compound has a high melting temperature due to its relatively high elastic constants.

**(d) Minimum Thermal Conductivity**

In any solid material, whether crystalline or amorphous, the minimum thermal conductivity, or $k_{min}$, carried by the atomic vibrations has major technological implications. A material's thermal conductivity reduces as temperature rises. The lowest limit of a crystal's lattice thermal conductivity above the Debye temperature, where the vibrational scattering rate is limited by lattice spacing, is described by the theoretical minimum thermal conductivity. Defects impact phonon transport over distances significantly larger than interatomic spacing, the minimum thermal conductivity does not depend on the presence of defects. The minimum thermal conductivity can be estimated from the following equation [110–112]:

$$k_{min} = k_B v_m \left(\frac{nN_A\rho}{M}\right)^{\frac{2}{3}} \qquad (39)$$

**Table 10** lists the calculated values of isotropic minimum thermal conductivity. The minimum thermal conductivity of the compound increases with pressure because the average sound velocity increases with rising pressure. The minimum thermal conductivity of the compound is low. Thus, the phonon thermal conductivity $La_3Ni_2O_7$ is also expected to be low.

**(e) Acoustic Impedance**

Acoustic impedance, represented by the symbol $Z$, controls how sound energy is transmitted through a material. Materials with a high acoustic impedance improve sound wave transmission. The energy reflected and transmitted when a sound wave collides with a surface depends on how different the two media's acoustic impedances are from one another. The acoustic impedance of a medium can be ascertained using the following formula provided that the density and shear modulus of the medium are known. Most of the sound will either be reflected or transmitted if the difference in impedance between the two materials is almost equal. Compounds with $Z = 0$ and $Z \to \infty$ are characteristic of perfect hard and soft surfaces, whereas compounds with $Z \to \infty$ are characteristic of ideal soft surfaces. Additionally, sound energy cannot travel through a perfectly soft surface. The difference in acoustic impedance between two media has wide-ranging applications in industries such as aerospace, medical ultrasound imaging, automotive, transducer design, acoustic sensing, and the manufacturing of musical instruments [113]. An appropriate level of softness on a surface also stops sound waves from traveling through it. The following formula has been used to determine a medium's acoustic impedance given its density and shear modulus [58,114]:

$$Z = \sqrt{\rho\, G} \qquad (40)$$

Acoustic impedance is measured in Rayl: 1 Rayl = $kgm^{-2}\, s^{-1}$ = 1 $Nsm^3$. The evaluated acoustic impedance for $La_3Ni_2O_7$ is listed in **Table 10**.

**(f) Grüneisen Parameter**

The Grüneisen parameter can be used to determine the strength of phonon-phonon interactions in a particular material. It is affected by temperature as well as pressure. There are five different Grüneisen constants: acoustic ($\gamma_a$), elastic ($\gamma_{el}$), lattice ($\gamma_l$), thermodynamic ($\gamma_d$), and electronic ($\gamma_e$). The thermodynamic Grüneisen constant $\gamma_d$ is in agreement with the values of $\gamma_a$ and $\gamma_{el}$ for the majority of metals, ionic, and molecular crystals. The difference between $\gamma_a$ and $\gamma_{el}$ is larger in most materials, but it is much smaller in rare earth metals. The expansion or contraction of a material as a result of heating its conduction electrons is described by the electronic Grüneisen constant ($\gamma_e$). The elastic Grüneisen constants $\gamma_{el}$ of the structure under consideration has been calculated from their relation with the Poisson's ratio:

$$\gamma_{el} = \frac{3(1+\nu)Y}{2(2-3\nu)} \qquad (41)$$

The value of $\gamma_{el}$ affects a number of physical processes, such as thermal expansion, thermal conductivity, the temperature dependence of elastic properties, and the attenuation of acoustic waves. It also evaluates the anharmonic effects in a crystal, such as thermal expansion, and the temperature dependence of phonon frequencies and their damping. The higher the value of $\gamma_{el}$, the greater the anharmonicity, resulting in a lower phonon thermal conductivity. The pressure dependent values of the Grüneisen parameter of La$_3$Ni$_2$O$_7$ are tabulated in **Table 10**. These values are typical for solids with medium level of anharmonicity. The pressure dependent trend demonstrates the similar behavior of several other physical processes as described in the preceding sections.

**Table 10.** Calculated mass density ($\rho$ in gm cm$^{-3}$), longitudinal, transverse, and mean sound velocities ($v_l$, $v_t$, and $v_m$ in m sec$^{-1}$), Debye temperature ($\theta_D$ in K), heat capacity per unit volume ($\rho C_\rho$ in JK$^{-1}$m$^{-3}$), melting temperature ($T_m$ in K), minimum thermal conductivity ($k_{min}$ in Wm$^{-1}$K$^{-1}$), acoustic impedance (Z in Rayl), and Grüneisen parameter ($\gamma_{el}$) of La$_3$Ni$_2$O$_7$ at different pressures.

| Pressure (GPa) | $\rho$ | $v_l$ | $v_t$ | $v_m$ | $\theta_D$ | $\rho C_\rho$ ($\times 10^6$) | $T_m$ | $k_{min}$ | Z ($\times 10^7$) | $\gamma_{el}$ |
|---|---|---|---|---|---|---|---|---|---|---|
| 30 | 7.90 | 7345.71 | 3649.90 | 4091.91 | 542.14 | 3.66 | 2356.82 | 1.12 | 2.88 | 2.05 |
| 32 | 7.96 | 7467.07 | 3754.35 | 4193.65 | 556.91 | 3.68 | 2322.14 | 1.13 | 2.99 | 1.98 |
| 34 | 8.01 | 7573.34 | 3806.00 | 4252.46 | 565.99 | 3.71 | 2401.71 | 1.18 | 3.05 | 1.98 |
| 36 | 8.07 | 7272.69 | 3522.72 | 3952.22 | 527.20 | 3.73 | 2235.20 | 1.10 | 2.84 | 2.13 |
| 38 | 8.11 | 7727.50 | 3679.96 | 4132.17 | 552.24 | 3.76 | 2486.64 | 1.15 | 2.98 | 2.13 |
| 40 | 8.16 | 7892.43 | 3957.88 | 4423.67 | 592.43 | 3.78 | 2598.95 | 1.24 | 3.23 | 1.98 |

**(g) Anisotropies in Sound Velocity**

The composition of a material determines the velocity of sound wave propagation (both longitudinal and transverse), independent of frequency and material dimensions. In a system, every atom vibrates in three different modes: two transverse and one longitudinal. Elastic anisotropy in a crystal is suggested by the anisotropic characteristics of sound velocities, and vice versa. Only specific crystallographic directions contain the pure longitudinal and transverse modes in the case of an anisotropic crystal. The modes of sound propagation in all other directions are either quasi-transverse or quasi-longitudinal waves. In an orthorhombic system, the [100], [010], and [001] directions exhibit distinct pure transverse and longitudinal modes. For orthorhombic symmetry, the acoustic velocities along these principal directions can be represented as [78]:

$$[100]v_l = \sqrt{\frac{C_{11}}{\rho}}, \quad [010]v_{t1} = \sqrt{\frac{C_{66}}{\rho}}, \quad [001]v_{t2}\sqrt{\frac{C_{55}}{\rho}} \qquad (42)$$

$$[010]v_l = \sqrt{\frac{C_{22}}{\rho}}, \quad [100]v_{t1} = \sqrt{\frac{C_{66}}{\rho}}, \quad [001]v_{t2}\sqrt{\frac{C_{44}}{\rho}} \qquad (43)$$

$$[001]v_l = \sqrt{\frac{C_{33}}{\rho}}, \quad [100]v_{t1} = \sqrt{\frac{C_{55}}{\rho}} \quad [010]v_{t2}\sqrt{\frac{C_{44}}{\rho}} \qquad (44)$$

Where $v_{t1}$ and $v_{t2}$ correspond to the first and second transverse modes, respectively. The computed acoustic wave velocities for La$_3$Ni$_2$O$_7$ are provided in **Table 10.** Anisotropic characteristics in sound velocities result from the elastic anisotropy present in a crystal. The longitudinal sound velocities along the [100], [010], and [001] directions are determined by $C_{11}$, $C_{22}$, and $C_{33}$, correspondingly, while the transverse modes are determined by $C_{44}$, $C_{55}$, and $C_{66}$.

**Table 10**. The longitudinal $v_l$ (in ms$^{-1}$) and transverse wave velocities ($v_{t1}$ and $v_{t2}$ in ms$^{-1}$) along [100], [010], and [001] directions in orthorhombic La$_3$Ni$_2$O$_7$ under various pressures.

| Pressure (GPa) | [100] | | | [010] | | | [001] | | |
|---|---|---|---|---|---|---|---|---|---|
| | $[100]v_l$ | $[010]v_{t1}$ | $[001]v_{t2}$ | $[010]v_l$ | $[100]v_{t1}$ | $[001]v_{t2}$ | $[001]v_l$ | $[100]v_{t1}$ | $[010]v_{t2}$ |
| 30 | 7242.81 | 3762.41 | 3415.90 | 7229.16 | 3762.41 | 3415.53 | 8006.09 | 3415.90 | 3415.53 |
| 32 | 7446.80 | 4628.54 | 3429.66 | 7455.98 | 4628.54 | 3461.56 | 7343.42 | 3429.66 | 3461.56 |
| 34 | 7586.62 | 4672.28 | 3509.40 | 7533.36 | 4672.28 | 3508.51 | 7437.46 | 3509.40 | 3508.51 |
| 36 | 7424.97 | 3983.39 | 3466.78 | 7428.30 | 3983.39 | 3467.50 | 6719.09 | 3466.78 | 3467.50 |
| 38 | 7927.69 | 4125.65 | 3681.02 | 7922.17 | 4125.65 | 3675.49 | 7043.64 | 3681.02 | 3675.49 |
| 40 | 7877.51 | 4830.33 | 3693.69 | 7876.65 | 4830.33 | 3693.69 | 7700.67 | 3693.69 | 3693.69 |

### 3.7 Optical Properties

The optical properties of a material are crucial to understand how light interacts with it and how it is scattered, absorbed, reflected, and transmitted through it. It also plays a pivotal role in predicting whether it can be used for optoelectronic and photovoltaic device applications or not. In this study, the various energy-dependent optical parameters, including the absorption coefficient $α(ω)$, reflectivity $R(ω)$, optical conductivity $σ(ω)$, refractive index $n(ω)$, dielectric function $ε(ω)$, and energy loss function $L(ω)$ of La$_3$Ni$_2$O$_7$ have been calculated for photon energies up to 35 eV under pressures at 30 GPa and 40 GPa and the outcomes are depicted in **Figure 20 – 21**. As the material under study is anisotropic, we have computed $α(ω)$, $R(ω)$, $σ(ω)$, $n(ω)$, $ε(ω)$, and $L(ω)$ for [100], [010] and [001] polarization directions of the electric field. Due to the metallic nature of the compounds being studied, Drude damping needs to be taken into account when analyzing the optical parameters. Thus, for the optical parameter analysis, we have used a Drude damping of 0.05 eV. The obtained results are discussed in the following sub-sections.

**(a) Absorption Coefficient**

The amount of photon energy that a material absorbs is determined by its absorption coefficient ($α$). The electronic nature of a material (metal, semiconductor, or insulator) can be inferred from the absorption spectra. Additionally, this parameter offers details on the optimal solar energy conversion efficiency. In **Figures 20** (a) and **21** (a), the absorption coefficient, $α(ω)$, of La$_3$Ni$_2$O$_7$ is displayed for pressures of 30 and 40 GPa. The data represent the behavior of $α(ω)$ when the electric field is polarized along the crystallographic directions [100], [010], and [001], showing how the material's response varies with these orientations under high-pressure conditions. The absorption spectrum is observed to start at zero photon energy, signifying the metallic nature of the compound [115]. **Figures 20** (a) and **21** (a) reveal that the absorption coefficient is significantly high in the energy ranges of approximately 8–13 eV and 17–32 eV, with the latter range exhibiting higher values. A noticeable dip in the absorption occurs around 14.5 eV, after which $α(ω)$ begins to rise again for higher photon energy. The peak position of the absorption coefficient differs between the [100], [010], and the [001] polarization. The positions of the peak value of the absorption coefficient for [100] and [010] polarization are nearly identical. The two peak energies for [100] and [010] are around 9.5 eV and 22.5 eV, whereas for [001] the peak energies are around 11.5 eV and 23 eV. This indicates optical anisotropy of the compounds. It is interesting to note that the material exhibits significant absorption in the ultraviolet region of the electromagnetic spectrum suggesting that $La_3Ni_2O_7$ is a good absorber of ultraviolet radiation. Across the three polarization directions, $α(ω)$ experiences a sharp decline near 32 eV, which closely corresponds to the location of the loss peak and marks the onset of collective charge (plasma) oscillations.

**(b) Energy Loss Function**

The loss function, $L(ω)$ estimates the energy losses experienced by electrons as they interact with the material due to the excitation of the plasma oscillation [116,117]. The calculated pressure-dependent $L(ω)$ of $La_3Ni_2O_7$ is shown in **Figures 20** (b) and **21** (b). This behavior can be conveniently characterized using the reciprocal of the imaginary part of the complex dielectric function. In other words, there is a negative correlation between the loss functions and the imaginary dielectric functions. When the peak of $L(ω)$ appears, the dielectric part approaches zero [117]. The energy loss spectra correspond to the frequency of collective oscillations of the electrons. The term bulk plasma frequency ($ω_p$) refers to the characteristic frequency/energy of the associated peaks, which are closely linked to the plasma oscillation [118]. From **Figures 20** (b) and **21** (b), we see that the peaks in $L(ω)$ occur at 32.6 eV and 33.1 eV for 30 and 40 GPa, respectively. Both reflectivity and absorption coefficient fall at the plasma frequency. This suggests that above the plasma energy, the compound under study turns transparent to electromagnetic waves [119].

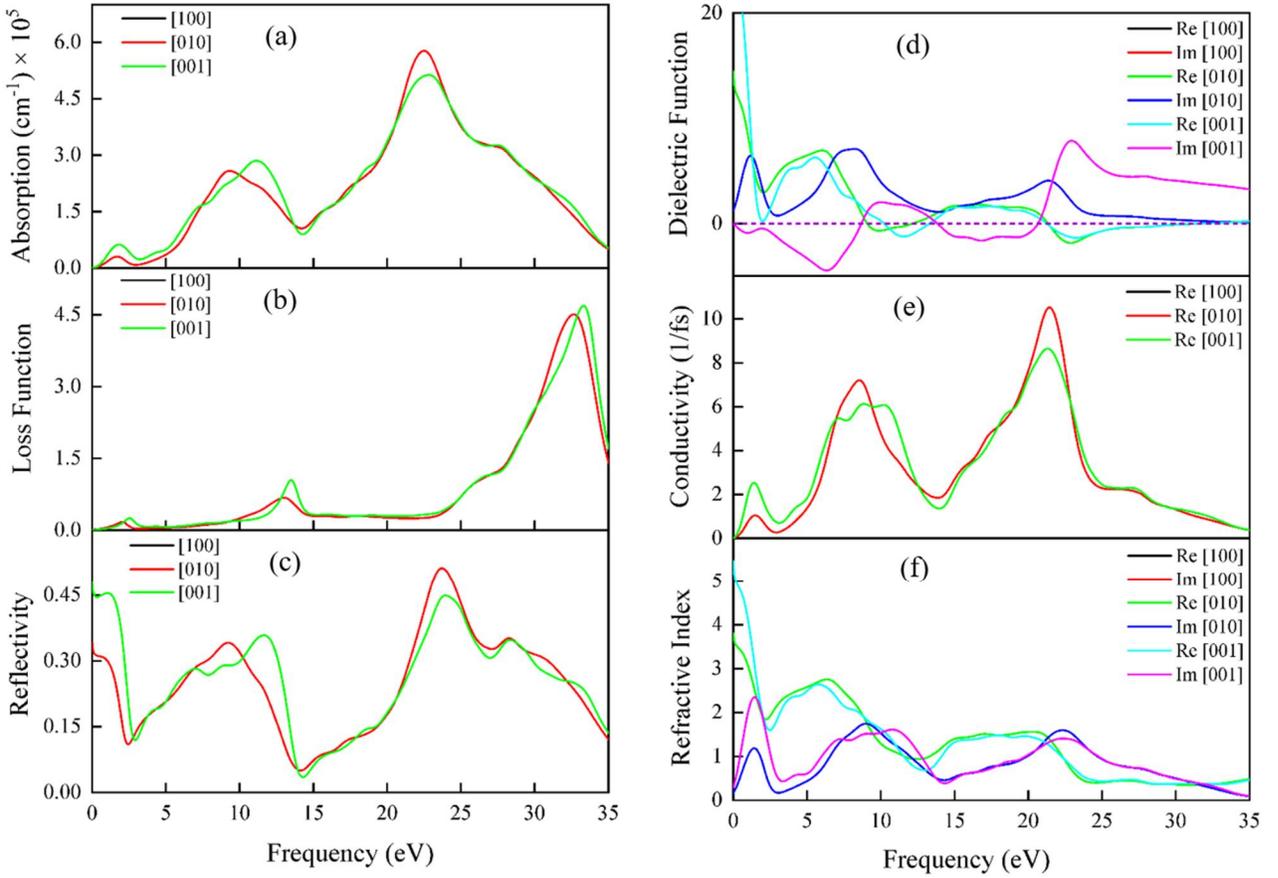

**Figure 20.** The energy (or, equivalently, frequency) dependent (a) absorption coefficient (b) loss function (c) reflectivity (d) dielectric function (e) optical conductivity, and (f) refractive index of $La_3Ni_2O_7$ with electric field polarization vectors along [100], [010], and [001] directions for 30 GPa pressure.

**(c) Reflectivity**

The computed frequency-dependent optical reflectivity ($R$) profiles of $La_3Ni_2O_7$ compounds along [100], [010], and [001] directions are depicted in **Figures 20** (c) and **21** (c). The reflectivity $R(\omega)$ starts with a value of ~ 32% at zero photon energy, decreases to ~ 11% (visible region), and rises to maximum values of ~ 34% (ultraviolet (UV) region) at around 9.4 eV. The reflectivity then decreases to a very low value of ~ 5% and rises to maximum values of ~ 51% (ultraviolet (UV) region) at around 24 eV and again starts to decrease for [100], and [010] polarizations. But for [001] direction, $R(\omega)$ starts with a value of ~ 45% at zero photon energy, decreases to ~ 12% (visible region), and rises to maximum values of ~ 36% (ultraviolet (UV) region) at around 11.5 eV. The reflectivity then decreases to a very low value of ~ 3.5% and rises to maximum values of ~ 45 % (ultraviolet (UV) region) at around 24 eV and again starts to decrease. The value of $R(\omega)$ then decreases drastically in the ultraviolet energy region. The reflectivity almost always stays below 35% throughout and, in some cases, it falls to a very low value. This suggests that the compound $La_3Ni_2O_7$ can be used as an antireflection system [120]. Optical anisotropy is moderate and the effect of pressure on the reflectivity is weak.

**(d) Dielectric Function**

The mechanism by which a material reacts to electromagnetic radiation in the infrared (IR)-visible to ultraviolet (UV) regions can be fully explained by the complex dielectric function, $\varepsilon(\omega)$. It can be used to calculate other optical constants, including optical conductivity, refractive index, reflectivity, energy loss function, and absorption of light. The dielectric function's real (Re) component, $\varepsilon_1(\omega)$ describes the material's electric polarization. On the other hand, the dielectric loss in the optical system is represented by the imaginary (Im) part, $\varepsilon_2(\omega)$ of the dielectric function. In metallic compounds, intraband electron transitions contribute to the optical spectra at lower energies (IR region), while interband transitions strongly depend on the electronic band structure [105,121]. **Figures 20** (d) and **21** (d) depict the calculated real and imaginary components of the dielectric function for $La_3Ni_2O_7$ at pressures of 30 and 40 GPa.

It is observed from **Figures 20** (d) and **21** (d) that the $\varepsilon_1(\omega)$ of $La_3Ni_2O_7$ crosses zero from below in the far UV range (about 30 to 32 eV), suggesting that the compound is metal. The values of $\varepsilon_2(\omega)$ reach zero around 34 eV across all polarization directions and both pressures, indicating that the material becomes transparent above this energy. In general, $\varepsilon_2$ becomes nonzero when absorption takes place. Additionally, the results indicate that effective plasma oscillations occur around 33 eV for all polarization directions. It is also noticed that the peak value of $\varepsilon_2(\omega)$ occurs in the lower photon energy (visible region). This suggests that the dielectric loss will be maximum at that peak energy. The dielectric spectra for [100] and [010] polarizations have nearly identical shapes which are different from [001] polarization direction. Both the real and imaginary parts of the dielectric function show small optical anisotropy, particularly at low energies whereas at high energies (above mid UV region) of the electromagnetic wave, the anisotropy diminishes dramatically. Once again, the effect pressure is very weak.

**(e) Optical Conductivity**

The conduction of free charge carriers within a specific range of photon energies is measured by optical conductivity, $\sigma(\omega)$, which is another crucial optoelectronic parameter of a material. **Figures 20** (e) and **21** (e) present the optical conductivity, $\sigma(\omega)$, of $La_3Ni_2O_7$ at pressures of 30 and 40 GPa for electric field polarizations in the [100], [010], and [001] directions. Photoconductivity begins at zero photon energy for all polarization directions, indicating that the material does not have a band gap in accordance with the band structure (**Figure 10**) and TDOS results (**Figure 12**). The spectra of all three polarizations exhibit a little variation in the range of photon energy from 0 to ~25 eV and $La_3Ni_2O_7$ has a higher peak in photoconductivity along [100], [010] polarizations compared to [001] polarizations and once again for [100] and [010] directions have a nearly identical shape. The effect pressure as well as anisotropy is very low. The anisotropy disappears in the far ultraviolet region (above ~27 eV) for all three polarization directions.

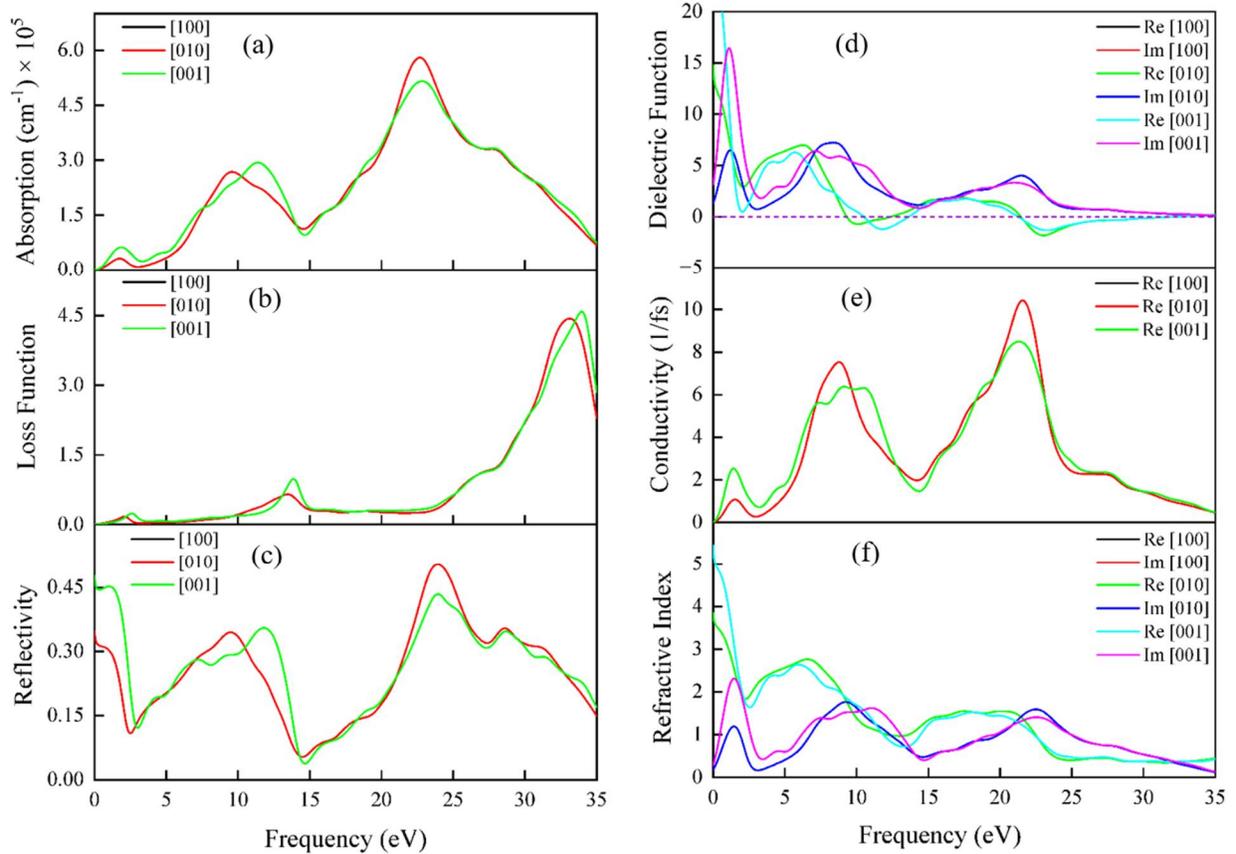

**Figure 21.** The energy (or, equivalently, frequency) dependent (a) absorption coefficient (b) loss function (c) reflectivity (d) dielectric function (e) optical conductivity, and (f) refractive index of $La_3Ni_2O_7$ with electric field polarization vectors along [100], [010], and [001] directions for 40 GPa.

**(f) Refractive Index**

The complex refractive index consists of two components: the real part, $n(\omega)$, and the imaginary part, $k(\omega)$. The real part of the refractive, $n(\omega)$ index analyzes the phase velocity of the photons in the medium, and the imaginary part or extinction coefficient, $k(\omega)$ elucidates the energy loss of the electromagnetic wave in the medium which is directly correlated to the dielectric constant and absorption coefficient [121]. The real and imaginary parts of the refractive index as a function of photon energy are illustrated in **Figures 20** (f) and **21** (f). The figures indicate that the real part of the refractive index is high in the infrared region and then gradually declines as it moves through the visible and ultraviolet regions. The spectra of all three polarizations exhibit a little variation in the photon energy range from 0 to ~20 eV. The *n (0)* (static) values of $La_3Ni_2O_7$ materials are higher for [001] direction (~5.10) compared to that for [100], [010] directions (~3.70). It can be concluded from the plots that $n(\omega)$ decreases sharply in the lower energy region (0 to ~2.70 eV) and then almost has the same value in the higher energy region. On the other hand, the extinction coefficient, $k(\omega)$ after reaching its maxima in the IR region, decreases in the UV region and then exhibits a little variation in spectra in the photon energy range from 0 to ~23 eV as seen in **Figures 20** (f) and **21** (f). In the IR and visible region, the refractive index and extinction coefficient are anisotropic. At higher energies, this anisotropy almost vanishes. The effect of pressure is weak.

The optical spectra presented in this section demonstrate a moderate level of anisotropy depending on the polarization state of the incident electromagnetic wave. Additionally, the degree of anisotropy is relatively greater at low energies. The impact of pressure is minimal, nearly negligible, across all optical spectra. This originates from the fact that pressure has a small effect on the electronic band structure within the range considered.

### 3.8 Superconducting State Properties

It has been reported that $La_3Ni_2O_7$ exhibits high transition temperature (high-$T_c$) superconductivity with a maximum $T_c$ of 80 K at pressures ranging from 14.0 to 43.5 GPa, as measured using high-pressure resistance and mutual inductive magnetic susceptibility techniques [18,93]. This is an example of superconductivity in a nickelate that is isostructural with high-$T_c$ cuprates [18,51,93,122–124] in some respect.

The superconducting $T_c$ of electron-phonon systems can be obtained from the widely used formula proposed by McMillan [125]:

$$T_C = \frac{\theta_D}{1.45} exp\left[\frac{-1.04(1 + \lambda_{ep})}{\lambda_{ep} - \mu^*(1 + 0.62\lambda_{ep})}\right] \qquad (45)$$

From this expression for critical temperature, it is evident that the superconducting transition temperature ($T_c$) depends on three fundamental physical parameters, specifically, the Debye temperature ($\theta_D$), the electron-phonon coupling constant ($\lambda_{ep}$), and the repulsive Coulomb pseudopotential ($\mu^*$). The pressure-dependent variation of DOS at the Fermi level and Coulomb pseudopotential is shown in **Figure 22**. For the pressure range considered, the pressure-dependent variation in the Debye temperature of $La_3Ni_2O_7$ is quite significant (**Table 10**). The electronic energy density of states at the Fermi level [$N(E_F)$], is nearly constant with rising pressure (**Figure 22** (a)). On the other hand, the density of states at the Fermi level can be used to estimate the repulsive Coulomb pseudopotential that diminishes $T_c$ [126]. The computed values of $\mu^*$ is ~ 0.195 (rounding off up to three decimal places) at all of our investigated pressures (**Figure 22** (b) and **Table 9**). This parameter is also a measure of electronic correlations. The computed values suggest that the strength of the electronic correlation is nearly constant within the pressure range considered for $La_3Ni_2O_7$.

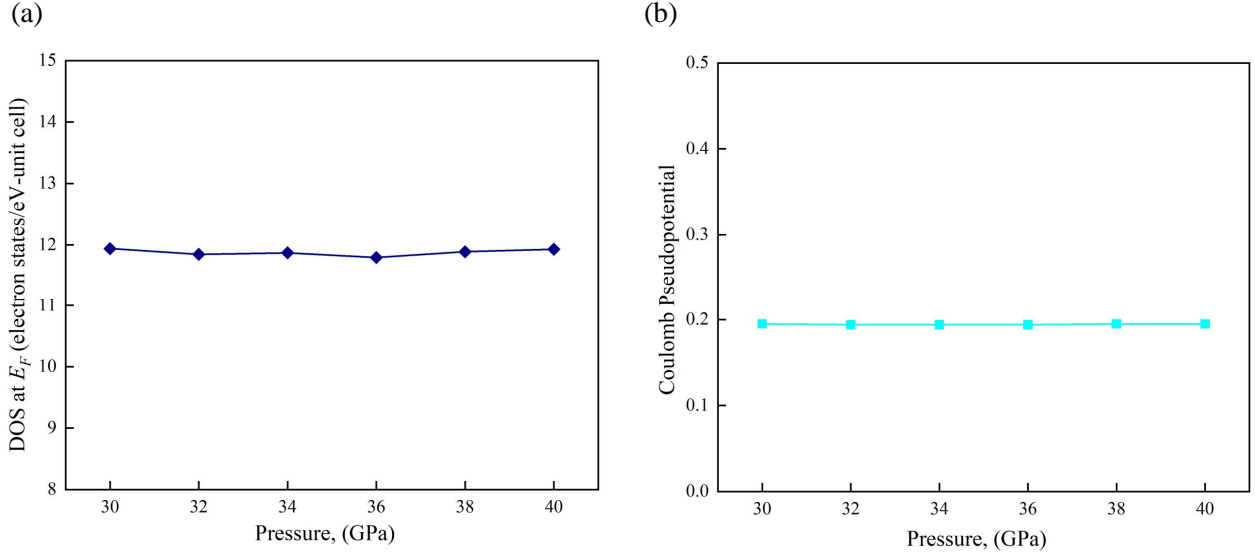

**Figure 22.** (a) DOS at $E_F$ and (b) Coulomb pseudopotential as a function of pressure for the compound $La_3Ni_2O_7$.

For full analytic calculation of the pressure dependence of $T_c$ within the electron-phonon scheme, information regarding $\lambda_{ep}$ is needed. Unfortunately, the CASTEP does not cover such calculations. At the same time, for a specific average electron-phonon interaction energy, $V_{ep}$, which is responsible for Fermi surface instability and the formation of Cooper pairs, $\lambda_{ep}$ varies with the value of $N(E_F)$ due to the relation, $\lambda_{ep} = N(E_F) V_{ep}$ [116]. This implies that $\lambda_{ep}$ might be nearly constant with increasing pressure for $La_3Ni_2O_7$ (**Figure 13**). Since $T_c$ strongly depends on the exponential (i.e., on $\lambda_{ep}$, and $\mu^*$) part of **Equation 45** which is roughly constant as $\lambda_{ep}$ and $\mu^*$ are nearly constant. Overall, the pressure dependent variation in the Debye temperature and nearly constant $\mu^*$ and $N(E_F)$ of $La_3Ni_2O_7$ should favor a slight enhancement of $T_c$ with rising pressure. Therefore, our work predicts a very weak pressure-dependent change in $T_c$ for $La_3Ni_2O_7$ within the pressure range considered. This prediction is quite compatible with the prior result [127]. But one important question remains; whether $La_3Ni_2O_7$ is an electron-phonon superconductor at all [127]. Further study is required to answer this vital question.

## 4. Conclusions

This study explores the pressure-dependent physical properties of the recently discovered nickelate superconductor, $La_3Ni_2O_7$ through a first-principles approach based on density functional theory. The structural parameters obtained from our calculations within the studied pressure range are consistent with previously reported results. The negative cohesive energy, along with mechanical and dynamical stability, suggests that $La_3Ni_2O_7$ is thermodynamically stable. $La_3Ni_2O_7$ exhibits ductility, excellent machinability, and a reasonable level of hardness within the studied pressure range. This combination of mechanical properties makes it highly suitable for device fabrication, while its high machinability index further highlights its potential for application as a dry lubricant. The analysis of various anisotropy indices shows that the compound displays a moderate level of elastic and mechanical anisotropy. The Poisson's ratio implies that central forces play a dominant role in atomic bonding. The electronic band structure and total density of states analyses suggest that the compound exhibits metallic behavior. Quite surprisingly, the electronic band structure showed negligible variation with pressure. The TDOS at the Fermi level remained mostly unaffected by pressure. Therefore, we expect the superconducting transition temperature of La$_3$Ni$_2$O$_7$, as a function of pressure, will be mainly governed by the pressure-driven variations in the Debye temperature or any other bosonic mode as appropriate. The primary contribution to the TDOS near the Fermi level comes from Ni 3$d$ and O 2$p$ states, with a smaller contribution from the La 6$s$ and La 5$d$ electronic states. The Fermi surfaces of La$_3$Ni$_2$O$_7$ exhibit both hole- and electron-like topologies and also indicate that electronic transport in this material is anisotropic. The calculated Debye temperatures are high for the compound across the studied pressure range, indicating its inherently hard nature. A clear correspondence among the Debye temperature ($\theta_D$), the melting temperature ($T_m$), and the minimum thermal conductivity ($k_{min}$) is observed. The optical spectra reveal metallic features and are in strong agreement with the TDOS profile. The characteristic peaks in reflectivity, refractive index, photoconductivity, and loss function show a slight shift to higher energies as pressure increases. The compound La$_3$Ni$_2$O$_7$ can be a useful antireflection system due to its fairly low reflectivity. The compound is also capable of efficiently absorbing ultraviolet radiation. The optical parameters exhibit moderate anisotropy and a minimal effect of pressure. We have explored the pressure dependence of the superconducting transition temperature in La$_3$Ni$_2$O$_7$ qualitatively. In the pressure range adopted, a small change in $T_c$ is predicted.

In summary, we have provided an in-depth analysis of the structural, elastic, electronic, optical, thermodynamic, and superconducting state properties of the recently discovered bilayer nickelate superconductor, La$_3$Ni$_2$O$_7$. The effect of pressure on various physical properties has been explored. Some novel results are reported. The compound under investigation exhibits several intriguing features, which have been discussed. We hope this study will encourage researchers to explore this fascinating material more thoroughly, both theoretically and experimentally in near future.

**Acknowledgments**

S.H.N. acknowledges the research grant (1151/5/52/RU/Science-07/19-20) from the Faculty of Science, University of Rajshahi, Bangladesh, which partly supported this work. M.E.H. gratefully acknowledges the National Science and Technology (NST) Fellowship, awarded by the Ministry of Science and Technology, Bangladesh, for supporting his M.Sc. research. This work is dedicated to the cherished memory of the martyrs of the July-August 2024 revolution in Bangladesh, whose sacrifices will forever inspire us.

**Data Availability**

The data sets generated and/or analyzed in this study are available from the corresponding author upon reasonable request.

**List of References**

[1] J. G. Bednorz and K. A. Müller, Possible high-$T_c$ superconductivity in the Ba-La-Cu-O system, Z. Physik B - Condensed Matter **64**, 189 (1986).

[2] V. I. Anisimov, D. Bukhvalov, and T. M. Rice, Electronic structure of possible nickelate analogs to the cuprates, Phys. Rev. B **59**, 7901 (1999).

[3] A. S. Botana, F. Bernardini, and A. Cano, Nickelate superconductors: An ongoing dialog between theory and experiments, Journal of Experimental and Theoretical Physics **132**, 618 (2021).

[4] D. Li, K. Lee, B. Y. Wang, M. Osada, S. Crossley, H. R. Lee, Y. Cui, Y. Hikita, and H. Y. Hwang, Superconductivity in an infinite-layer nickelate, Nature **572**, 624 (2019).

[5] M. Keenari, Study of Reduction Process on Perovskite Nickelates and Its Derivatives: A Bulk and Thin Film Approach, PhD Thesis, Normandie Université, 2023.

[6] S. Larsson, Conductivity Properties of Perovskite Nickelates and Cuprates Depend on the Oxidation States of the Metal Ions, J Supercond Nov Magn **35**, 3101 (2022).

[7] G. Zhou et al., *Gigantic-Oxidative Atomic-Layer-by-Layer Epitaxy for Artificially Designed Complex Oxides*, arXiv:2406.16520.

[8] F. Lechermann, Late transition metal oxides with infinite-layer structure: Nickelates versus cuprates, Phys. Rev. B **101**, 081110 (2020).

[9] M. Kitatani, L. Si, P. Worm, J. M. Tomczak, R. Arita, and K. Held, Optimizing Superconductivity: From Cuprates via Nickelates to Palladates, Phys. Rev. Lett. **130**, 166002 (2023).

[10] K. Held, L. Si, P. Worm, O. Janson, R. Arita, Z. Zhong, J. M. Tomczak, and M. Kitatani, Phase diagram of nickelate superconductors calculated by dynamical vertex approximation, Frontiers in Physics **9**, 810394 (2022).

[11] C. Lane, J. W. Furness, I. G. Buda, Y. Zhang, R. S. Markiewicz, B. Barbiellini, J. Sun, and A. Bansil, Antiferromagnetic ground state of $La_2CuO_4$: A parameter-free *ab initio* description, Phys. Rev. B **98**, 125140 (2018).

[12] B. Szpunar, V. H. Smith Jr, and R. W. Smith, Electronic structure of antiferromagnetic $YBa_2Cu_3O_6$, Physica C: Superconductivity **152**, 91 (1988).

[13] M. Kitatani, L. Si, O. Janson, R. Arita, Z. Zhong, and K. Held, Nickelate superconductors—a renaissance of the one-band Hubbard model, Npj Quantum Materials **5**, 59 (2020).

[14] M. Osada, B. Y. Wang, K. Lee, D. Li, and H. Y. Hwang, Phase diagram of infinite layer praseodymium nickelate $Pr_{1-x}Sr_xNiO_2$ thin films, Phys. Rev. Materials **4**, 121801 (2020).

[15] K. V. Mitsen and O. M. Ivanenko, Superconducting phase diagrams of cuprates and pnictides as a key to understanding the HTSC mechanism, Physics-Uspekhi **60**, 402 (2017).

[16] Y. Zhang, D. Su, Y. Huang, Z. Shan, H. Sun, M. Huo, K. Ye, J. Zhang, Z. Yang, and Y. Xu, High-temperature superconductivity with zero resistance and strange-metal behaviour in $La_3Ni_2O_{7-\delta}$, Nature Physics 1 (2024).

[17] C. D. Ling, D. N. Argyriou, G. Wu, and J. J. Neumeier, Neutron diffraction study of $La_3Ni_2O_7$: Structural relationships among n= 1, 2, and 3 phases $La_{n+1}Ni_nO_{3n+1}$, Journal of Solid State Chemistry **152**, 517 (2000).

[18] H. Sun et al., Superconductivity near 80 Kelvin in single crystals of $La_3Ni_2O_7$ under pressure, Nature **621**, 493 (2023).

[19] D. Li, K. Lee, B. Y. Wang, M. Osada, S. Crossley, H. R. Lee, Y. Cui, Y. Hikita, and H. Y. Hwang, Superconductivity in an infinite-layer nickelate, Nature **572**, 624 (2019).

[20] F. Lechermann, J. Gondolf, S. Bötzel, and I. M. Eremin, Electronic correlations and superconducting instability in $La_3Ni_2O_7$ under high pressure, Phys. Rev. B **108**, L201121 (2023).

[21] Y. A. O. Daoxin, Theoretical study of $La_3Ni_2O_7$ and $La_4Ni_3O_{10}$, Science & Technology Review 1 (2024).

[22] Z. Fan, J.-F. Zhang, B. Zhan, D. Lv, X.-Y. Jiang, B. Normand, and T. Xiang, Superconductivity in nickelate and cuprate superconductors with strong bilayer coupling, Phys. Rev. B **110**, 024514 (2024).

[23] D.-C. Lu, M. Li, Z.-Y. Zeng, W. Hou, J. Wang, F. Yang, and Y.-Z. You, *Superconductivity from Doping Symmetric Mass Generation Insulators: Application to* $La_3Ni_2O_7$ *under Pressure*, arXiv:2308.11195.

[24] M. Q. Cai, G. W. Yang, X. Tan, Y. L. Cao, L. L. Wang, W. Y. Hu, and Y. G. Wang, First-principles study of pressure-induced metal-insulator transition in $BiNiO_3$, Applied Physics Letters **91**, (2007).

[25] N. H. Jo, L. Xiang, U. S. Kaluarachchi, M. Masters, K. Neilson, S. S. Downing, P. C. Canfield, and S. L. Bud'ko, Pressure induced change in the electronic state of $Ta_4Pd_3Te_{16}$, Phys. Rev. B **95**, 134516 (2017).

[26] S. H. Naqib, M. T. Hoque, and A. Islam, *Oxygen Depletion Dependence of Pressure Coefficient of YBCO (123)*, in *Advances in High Pressure Science and Technology: Proceedings of the Fourth National Conference on High Pressure Science and Technology* (1997).

[27] R. S. Islam, S. H. Naqib, and A. K. M. A. Islam, *LATTICE GAS PHENOMENOLOGY, VAN HOVE SCENARIO AND THE COMPLEX DOPING DEPENDENCE OF $dT_c / dP$ OF $YBa_2 Cu_3 O$*

$_{6+x}$, in *Magnetic and Superconducting Materials* (World Scientific Publishing Company, Tehran, Iran, 2000), pp. 91–98.

[28] M. I. Naher and S. H. Naqib, Structural, elastic, electronic, bonding, and optical properties of topological CaSn$_3$ semimetal, Journal of Alloys and Compounds **829**, 154509 (2020).

[29] B. R. Rano, I. M. Syed, and S. H. Naqib, Elastic, electronic, bonding, and optical properties of WTe$_2$ Weyl semimetal: A comparative investigation with MoTe$_2$ from first principles, Results in Physics **19**, 103639 (2020).

[30] F. Parvin and S. H. Naqib, Structural, elastic, electronic, thermodynamic, and optical properties of layered BaPd$_2$As$_2$ pnictide superconductor: A first principles investigation, Journal of Alloys and Compounds **780**, 452 (2019).

[31] Y. Al-Douri, M. Ameri, A. Bouhemadou, and K. M. Batoo, First-Principles Calculations to Investigate the Refractive Index and Optical Dielectric Constant of Na$_3$Sb$X_4$ ($X$ = S, Se) Ternary Chalcogenides, Physica Status Solidi (b) **256**, 1900131 (2019).

[32] I. E. Yahiaoui, A. Lazreg, Z. Dridi, Y. Al-douri, and B. Bouhafs, Gd impurities effect on alloy: first-principle calculations, Bull Mater Sci **41**, 2 (2018).

[33] W. Kohn and L. J. Sham, Self-Consistent Equations Including Exchange and Correlation Effects, Phys. Rev. **140**, A1133 (1965).

[34] J. P. Perdew, K. Burke, and M. Ernzerhof, Generalized Gradient Approximation Made Simple, Phys. Rev. Lett. **77**, 3865 (1996).

[35] S. J. Clark, M. D. Segall, C. J. Pickard, P. J. Hasnip, M. I. J. Probert, K. Refson, and M. C. Payne, First principles methods using CASTEP, Zeitschrift Für Kristallographie - Crystalline Materials **220**, 567 (2005).

[36] M. Lewin, E. H. Lieb, and R. Seiringer, The local density approximation in density functional theory, Pure and Applied Analysis **2**, 35 (2019).

[37] V. Sahni, K.-P. Bohnen, and M. K. Harbola, Analysis of the local-density approximation of density-functional theory, Phys. Rev. A **37**, 1895 (1988).

[38] *General Methods for Geometry and Wave Function Optimization | The Journal of Physical Chemistry*, https://pubs.acs.org/doi/abs/10.1021/j100203a036.

[39] *Phys. Rev. B 13, 5188 (1976) - Special Points for Brillouin-Zone Integrations*, https://journals.aps.org/prb/abstract/10.1103/PhysRevB.13.5188.

[40] O. H. Nielsen and R. M. Martin, First-Principles Calculation of Stress, Phys. Rev. Lett. **50**, 697 (1983).

[41] J. P. Watt, Hashin-Shtrikman bounds on the effective elastic moduli of polycrystals with orthorhombic symmetry, Journal of Applied Physics **50**, 6290 (1979).

[42] J. P. Watt and L. Peselnick, Clarification of the Hashin-Shtrikman bounds on the effective elastic moduli of polycrystals with hexagonal, trigonal, and tetragonal symmetries, Journal of Applied Physics **51**, 1525 (1980).

[43] F. Parvin and S. Naqib, Pressure dependence of structural, elastic, electronic, thermodynamic, and optical properties of van der Waals-type $NaSn_2P_2$ pnictide superconductor: Insights from DFT study, Results in Physics **21**, (2021).

[44] S. Hadji, A. Bouhemadou, K. Haddadi, D. Cherrad, R. Khenata, S. Bin-Omran, and Y. Al-Douri, Elastic, electronic, optical and thermodynamic properties of $Ba_3Ca_2Si_2N_6$ semiconductor: First-principles predictions, Physica B: Condensed Matter **589**, 412213 (2020).

[45] S. Touam et al., First-principles computations of As-ternary alloys: a study on structural, electronic, optical and elastic properties, Bull Mater Sci **43**, 22 (2020).

[46] A. Chowdhury, M. A. Ali, M. M. Hossain, M. M. Uddin, S. H. Naqib, and A. K. M. A. Islam, Predicted MAX Phase $Sc_2InC$: Dynamical Stability, Vibrational and Optical Properties, Physica Status Solidi (b) **255**, 1700235 (2018).

[47] M. Roknuzzaman, M. A. Hadi, M. J. Abden, M. T. Nasir, A. K. M. A. Islam, M. S. Ali, K. Ostrikov, and S. H. Naqib, Physical properties of predicted $Ti_2CdN$ versus existing Ti2CdC MAX phase: An *ab initio* study, Computational Materials Science **113**, 148 (2016).

[48] M. M. Hossain and S. H. Naqib, Structural, elastic, electronic, andoptical properties of layered TiNX (X = F, Cl, Br, I) compounds: a density functional theory study, Molecular Physics **118**, e1609706 (2020).

[49] M. S. Islam, R. Ahmed, M. Mahamudujjaman, R. S. Islam, and S. H. Naqib, A comparative study of the structural, elastic, thermophysical, and optoelectronic properties of $CaZn_2X_2$ (X= N, P, As) semiconductors via ab-initio approach, Results in Physics **44**, 106214 (2023).

[50] F. Birch, Finite strain isotherm and velocities for single-crystal and polycrystalline NaCl at high pressures and 300°K, Journal of Geophysical Research: Solid Earth **83**, 1257 (1978).

[51] K. Jiang, Z. Wang, and F.-C. Zhang, High-temperature superconductivity in $La_3Ni_2O_7$, Chinese Physics Letters **41**, 017402 (2024).

[52] *Iu ZTY, Zhou X, Khare SV, Gall D. J Phys Condens*

[53] Z. T. Y. Liu, D. Gall, and S. V. Khare, Electronic and bonding analysis of hardness in pyrite-type transition-metal pernitrides, Phys. Rev. B **90**, 134102 (2014).

[54] R. Golesorkhtabar, P. Pavone, J. Spitaler, P. Puschnig, and C. Draxl, ElaStic: A tool for calculating second-order elastic constants from first principles, Computer Physics Communications **184**, 1861 (2013).

[55] D. C. Wallace and H. Callen, Thermodynamics of crystals, American Journal of Physics **40**, 1718 (1972).

[56] C. Chen, L. Liu, Y. Wen, Y. Jiang, and L. Chen, Elastic properties of orthorhombic $YBa_2Cu_3O_7$ under pressure, Crystals **9**, 497 (2019).

[57] W. Voigt, *Lehrbuch Der Kristallphysik:(Mit Ausschluss Der Kristalloptik)*, Vol. 34 (BG Teubner, 1910).

[58] R. Hill, The elastic behaviour of a crystalline aggregate, Proceedings of the Physical Society. Section A **65**, 349 (1952).

[59] M. Jamal, S. J. Asadabadi, I. Ahmad, and H. R. Aliabad, Elastic constants of cubic crystals, Computational Materials Science **95**, 592 (2014).

[60] A. Gueddouh, B. Bentria, and I. K. Lefkaier, First-principle investigations of structure, elastic and bond hardness of $Fe_xB$ (x= 1, 2, 3) under pressure, Journal of Magnetism and Magnetic Materials **406**, 192 (2016).

[61] M. I. Naher and S. H. Naqib, A comprehensive study of the thermophysical and optoelectronic properties of $Nb_2P_5$ via ab-initio technique, Results in Physics **28**, 104623 (2021).

[62] P. Ravindran, L. Fast, P. A. Korzhavyi, B. Johansson, J. Wills, and O. Eriksson, Density functional theory for calculation of elastic properties of orthorhombic crystals: Application to $TiSi_2$, Journal of Applied Physics **84**, 4891 (1998).

[63] M. Rajagopalan, S. P. Kumar, and R. Anuthama, FP-LAPW study of the elastic properties of $Al_2X$ (X= Sc, Y, La, Lu), Physica B: Condensed Matter **405**, 1817 (2010).

[64] C. Kittel, Introduction to solid state physics, eight editions, library of congress cataloging, (2005).

[65] A. Yildirim, H. Koc, and E. Deligoz, First-principles study of the structural, elastic, electronic, optical, and vibrational properties of intermetallic $Pd_2Ga$, Chinese Physics B **21**, 037101 (2012).

[66] W. Kim, Strategies for engineering phonon transport in thermoelectrics, Journal of Materials Chemistry C **3**, 10336 (2015).

[67] S. F. Pugh, XCII. Relations between the elastic moduli and the plastic properties of polycrystalline pure metals, The London, Edinburgh, and Dublin Philosophical Magazine and Journal of Science **45**, 823 (1954).

[68] M. M. Hossain, M. A. Ali, M. M. Uddin, A. Islam, and S. H. Naqib, Origin of high hardness and optoelectronic and thermo-physical properties of boron-rich compounds $B_6X$ (X= S, Se): a comprehensive study via DFT approach, Journal of Applied Physics **129**, (2021).

[69] Z. Yang, D. Shi, B. Wen, R. Melnik, S. Yao, and T. Li, First-principle studies of ca–x (x= si, ge, sn, pb) intermetallic compounds, Journal of Solid State Chemistry **183**, 136 (2010).

[70] M. L. Ali, M. M. Billah, M. Khan, M. N. M. Nobin, and M. Z. Rahaman, Pressure-induced physical properties of alkali metal chlorides $Rb_2NbCl_6$: A density functional theory study, AIP Advances **13**, (2023).

[71] M. Mattesini, R. Ahuja, and B. Johansson, Cubic $Hf_3N_4$ and $Zr_3N_4$: A class of hard materials, Physical Review B **68**, 184108 (2003).

[72] P. H. Mott, J. R. Dorgan, and C. M. Roland, The bulk modulus and Poisson's ratio of "incompressible" materials, Journal of Sound and Vibration **312**, 572 (2008).

[73] H. Fu, D. Li, F. Peng, T. Gao, and X. Cheng, Ab initio calculations of elastic constants and thermodynamic properties of NiAl under high pressures, Computational Materials Science **44**, 774 (2008).

[74] A. Šim\uunek, How to estimate hardness of crystals on a pocket calculator, Physical Review B—Condensed Matter and Materials Physics **75**, 172108 (2007).

[75] W. Feng and S. Cui, Mechanical and electronic properties of $Ti_2AlN$ and $Ti_4AlN_3$: a first-principles study, Canadian Journal of Physics **92**, 1652 (2014).

[76] D. G. Pettifor, Theoretical predictions of structure and related properties of intermetallics, Materials Science and Technology **8**, 345 (1992).

[77] L. Kleinman, Deformation Potentials in Silicon. I. Uniaxial Strain, Phys. Rev. **128**, 2614 (1962).

[78] Z. Sun, D. Music, R. Ahuja, and J. M. Schneider, Theoretical investigation of the bonding and elastic properties of nanolayered ternary nitrides, Phys. Rev. B **71**, 193402 (2005).

[79] M. A. Hadi, S.-R. Christopoulos, S. H. Naqib, A. Chroneos, M. E. Fitzpatrick, and A. Islam, Physical properties and defect processes of $M_3SnC_2$ (M= Ti, Zr, Hf) MAX phases: Effect of M-elements, Journal of Alloys and Compounds **748**, 804 (2018).

[80] A. Chowdhury, M. A. Ali, M. M. Hossain, M. M. Uddin, S. H. Naqib, and A. K. M. A. Islam, Predicted MAX Phase $Sc_2InC$: Dynamical Stability, Vibrational and Optical Properties, Physica Status Solidi (b) **255**, 1700235 (2018).

[81] M. A. Ali, M. T. Nasir, M. R. Khatun, A. Islam, and S. H. Naqib, An ab initio investigation of vibrational, thermodynamic, and optical properties of $Sc_2AlC$ MAX compound, Chinese Physics B **25**, 103102 (2016).

[82] W. A. Harrison, Instructor's guide and solutions manual for Electronic structure and the properties of solids: the physics of the chemical bond, (No Title) (1980).

[83] M. I. Naher, F. Parvin, A. K. M. A. Islam, and S. H. Naqib, Physical properties of niobium-based intermetallics ($Nb_3B$; B = Os, Pt, Au): a DFT-based ab-initio study, Eur. Phys. J. B **91**, 289 (2018).

[84] X. Gao, Y. Jiang, R. Zhou, and J. Feng, Stability and elastic properties of Y–C binary compounds investigated by first principles calculations, Journal of Alloys and Compounds **587**, 819 (2014).

[85] A. Tasnim, M. Mahamudujjaman, M. A. Afzal, R. S. Islam, and S. H. Naqib, Pressure-dependent semiconductor–metal transition and elastic, electronic, optical, and thermophysical properties of orthorhombic SnS binary chalcogenide, Results in Physics **45**, 106236 (2023).

[86] C. M. Kube and M. De Jong, Elastic constants of polycrystals with generally anisotropic crystals, Journal of Applied Physics **120**, (2016).

[87] S. I. Ranganathan and M. Ostoja-Starzewski, Universal Elastic Anisotropy Index, Phys. Rev. Lett. **101**, 055504 (2008).

[88] F. Vahldiek, *Anisotropy in Single-Crystal Refractory Compounds* (Springer, 2013).

[89] D. H. Chung and W. R. Buessem, The elastic anisotropy of crystals, Journal of Applied Physics **38**, 2010 (1967).

[90] V. Arsigny, P. Fillard, X. Pennec, and N. Ayache, *Fast and Simple Calculus on Tensors in the Log-Euclidean Framework*, in *Medical Image Computing and Computer-Assisted Intervention – MICCAI 2005*, edited by J. S. Duncan and G. Gerig, Vol. 3749 (Springer Berlin Heidelberg, Berlin, Heidelberg, 2005), pp. 115–122.

[91] M. I. Naher and S. H. Naqib, An ab-initio study on structural, elastic, electronic, bonding, thermal, and optical properties of topological Weyl semimetal Ta X (X= P, As), Scientific Reports **11**, 5592 (2021).

[92] R. Gaillac, P. Pullumbi, and F.-X. Coudert, ELATE: an open-source online application for analysis and visualization of elastic tensors, Journal of Physics: Condensed Matter **28**, 275201 (2016).

[93] J. Hou, P.-T. Yang, Z.-Y. Liu, J.-Y. Li, P.-F. Shan, L. Ma, G. Wang, N.-N. Wang, H.-Z. Guo, and J.-P. Sun, Emergence of high-temperature superconducting phase in pressurized $La_3Ni_2O_7$ crystals, Chinese Physics Letters **40**, 117302 (2023).

[94] K. Boudiaf, A. Bouhemadou, Y. Al-Douri, R. Khenata, S. Bin-Omran, and N. Guechi, Electronic and thermoelectric properties of the layered BaFAgCh (Ch= S, Se and Te): first-principles study, Journal of Alloys and Compounds **759**, 32 (2018).

[95] A. Bekhti-Siad, K. Bettine, D. P. Rai, Y. Al-Douri, X. Wang, R. Khenata, A. Bouhemadou, and C. H. Voon, Electronic, optical and thermoelectric investigations of Zintl phase $AE_3AlAs_3$ (AE= Sr, Ba): first-principles calculations, Chinese Journal of Physics **56**, 870 (2018).

[96] A. Belhachemi, H. Abid, Y. Al-Douri, M. Sehil, A. Bouhemadou, and M. Ameri, First-principles calculations to investigate the structural, electronic and optical properties of $Zn_{1-x}Mg_xTe$ ternary alloys, Chinese Journal of Physics **55**, 1018 (2017).

[97] K. H. Bennemann and J. W. Garland, *Theory for Superconductivity in D-Band Metals*, in *AIP Conference Proceedings*, Vol. 4 (American Institute of Physics, 1972), pp. 103–137.

[98] B. L. Gyorffy, *A Theory of the Electron-Phonon Interaction and the Superconducting Transition Temperature, Tc, in Strongly Scattering Systems*, in *Superconductivity in D- and f-Band Metals*, edited by D. H. Douglass (Springer US, Boston, MA, 1976), pp. 29–57.

[99] J. P. Carbotte, Properties of boson-exchange superconductors, Rev. Mod. Phys. **62**, 1027 (1990).

[100] M. M. Mridha and S. H. Naqib, Pressure dependent elastic, electronic, superconducting, and optical properties of ternary barium phosphides ($BaM_2P_2$; M= Ni, Rh): DFT based insights, Physica Scripta **95**, 105809 (2020).

[101] N. E. Christensen and D. L. Novikov, Calculated superconductive properties of Li and Na under presure, Phys. Rev. B, 224508 **73**, 1 (2006).

[102] J. Bardeen, L. N. Cooper, and J. R. Schrieffer, Theory of Superconductivity, Phys. Rev. **108**, 1175 (1957).

[103] G. Grimvall and S. Sjödin, Correlation of properties of materials to Debye and melting temperatures, Physica Scripta **10**, 340 (1974).

[104] Y. Benkaddour et al., First-Principle Calculations of Structural, Elastic, and Electronic Properties of Intermetallic Rare Earth $R_2Ni_2Pb$ (R = Ho, Lu, and Sm) Compounds, J Supercond Nov Magn **31**, 395 (2018).

[105] F. Parvin and S. H. Naqib, Elastic, thermodynamic, electronic, and optical properties of recently discovered superconducting transition metal boride NbRuB: An ab-initio investigation, Chinese Physics B **26**, 106201 (2017).

[106] O. L. Anderson, A simplified method for calculating the Debye temperature from elastic constants, Journal of Physics and Chemistry of Solids **24**, 909 (1963).

[107] X.-W. Sun, N. Bioud, Z.-J. Fu, X.-P. Wei, T. Song, and Z.-W. Li, High-pressure elastic properties of cubic $Ir_2P$ from ab initio calculations, Physics Letters A **380**, 3672 (2016).

[108] A. Alam, F. Parvin, and S. H. Naqib, First-principles pressure dependent investigation of the physical properties of $KB_2H_8$: a prospective high-$T_C$ superconductor, Results in Physics **58**, 107498 (2024).

[109] M. E. Fine, L. D. Brown, and H. L. Marcus, Elastic constants versus melting temperature in metals, Scripta Metallurgica **18**, 951 (1984).

[110] D. R. Clarke, Materials selection guidelines for low thermal conductivity thermal barrier coatings, Surface and Coatings Technology **163**, 67 (2003).

[111] M. M. Hossain, M. A. Ali, M. M. Uddin, M. A. Hossain, M. Rasadujjaman, S. H. Naqib, M. Nagao, S. Watauchi, and I. Tanaka, Influence of Se doping on recently synthesized $NaInS_{2-x}Se_x$ solid solutions for potential thermo-mechanical applications studied via first-principles method, Materials Today Communications **26**, 101988 (2021).

[112] M. M. Hossain, M. A. Hossain, S. A. Moon, M. A. Ali, M. M. Uddin, S. H. Naqib, A. Islam, M. Nagao, S. Watauchi, and I. Tanaka, $NaInX_2$ (X= S, Se) layered materials for energy harvesting applications: first-principles insights into optoelectronic and thermoelectric properties, Journal of Materials Science: Materials in Electronics **32**, 3878 (2021).

[113] M. I. Naher and S. H. Naqib, Possible applications of $Mo_2C$ in the orthorhombic and hexagonal phases explored via ab-initio investigations of elastic, bonding, optoelectronic and thermophysical properties, Results in Physics **37**, 105505 (2022).

[114] I. N. Frantsevich, F. F. Voronov, and S. A. Bakuta, Elastic constants and elastic moduli of metals and nonmetals (In Russian), Kiev, Izdatel'stvo Naukova Dumka, 1982, 288 (1982).

[115] F. Sultana, M. M. Uddin, M. A. Ali, M. M. Hossain, S. H. Naqib, and A. Islam, First principles study of $M_2InC$ (M= Zr, Hf and Ta) MAX phases: the effect of M atomic species, Results in Physics **11**, 869 (2018).

[116] M. A. H. Shah, M. I. Naher, and S. H. Naqib, *First-Principles Exploration of the Pressure Dependent Physical Properties of $Sn_4Au$: A Superconducting Topological Semimetal*, arXiv:2408.07451.

[117] M. Rizwan, H. F. Arooj, F. Noor, K. Nawaz, M. A. Ullah, Z. Usman, A. Akremi, and T. Mahmood, Computational study to investigate effectiveness of titanium substitution in $CaFeH_3$ perovskite-type hydride: an approach towards advanced hydrogen storage system, Journal of Materials Research and Technology **31**, 2676 (2024).

[118] S. Azad, B. R. Rano, I. M. Syed, and S. H. Naqib, A comparative study of the physical properties of layered transition metal nitride halides MNCl (M= Zr, Hf): DFT based insights, Physica Scripta **98**, 115982 (2023).

[119] M. Roknuzzaman, M. A. Hadi, M. A. Ali, M. M. Hossain, N. Jahan, M. M. Uddin, J. A. Alarco, and K. Ostrikov, First hafnium-based MAX phase in the 312 family, $Hf_3AlC_2$: A first-principles study, Journal of Alloys and Compounds **727**, 616 (2017).

[120] S. Li, R. Ahuja, M. W. Barsoum, P. Jena, and B. Johansson, Optical properties of $Ti_3SiC_2$ and $Ti_4AlN_3$, Applied Physics Letters **92**, (2008).

[121] D. Qu, L. Bao, Z. Kong, and Y. Duan, First-principles predictions of electronic, elastic, and optical properties of ScBC and YBC ternary cermet phases, Vacuum **179**, 109488 (2020).

[122] M. Wang, H.-H. Wen, T. Wu, D.-X. Yao, and T. Xiang, Normal and superconducting properties of $La_3Ni_2O_7$, Chinese Physics Letters **41**, 077402 (2024).

[123] Y. Zhang, L.-F. Lin, W. Hu, A. Moreo, S. Dong, and E. Dagotto, Similarities and differences between nickelate and cuprate films grown on a $SrTiO_3$ substrate, Phys. Rev. B **102**, 195117 (2020).

[124] R. Zhang, C. Lane, B. Singh, J. Nokelainen, B. Barbiellini, R. S. Markiewicz, A. Bansil, and J. Sun, Magnetic and f-electron effects in $LaNiO_2$ and $NdNiO_2$ nickelates with cuprate-like 3 $d_{x^2-y^2}$ band, Communications Physics **4**, 118 (2021).

[125] W. L. McMillan, Transition Temperature of Strong-Coupled Superconductors, Phys. Rev. **167**, 331 (1968).

[126] D. H. Douglass, Superconductivity in *d*-and *f*-Band Metals, (No Title) (1972).

[127] H. Sun, M. Huo, X. Hu, J. Li, Z. Liu, Y. Han, L. Tang, Z. Mao, P. Yang, and B. Wang, Signatures of superconductivity near 80 K in a nickelate under high pressure, Nature **621**, 493 (2023).

**Author Contributions**

**Md. Enamul Haque**: Methodology, Software, Writing- Original draft. **Ruman Ali**: Methodology, Software. **M. A. Masum**: Methodology, Software. **Jahid Hassan**: Methodology, Software. **S. H. Naqib**: Conceptualization, Supervision, Formal Analysis, Writing- Reviewing and Editing.

**Competing Interests**

The authors declare no competing interests.